\documentclass[prd,twocolumn,showpacs]{revtex4}
\usepackage{hyperref,amssymb,amsmath,mathrsfs,bm,graphicx}
\usepackage{enumitem}
\usepackage{color}
\begin{document}
\title{Axially symmetric static sources of gravitational field} 
\author{J. L. Hernandez-Pastora}
\email{jlhp@usal.es}
\affiliation{Departamento de Matematica Aplicada and Instituto Universitario de
Fisica Fundamental y Matematicas, Universidad de Salamanca, Salamanca, Spain}
\author{L. Herrera}
\email{lherrera@usal.es}
\affiliation{Escuela de F\'\i sica, Facultad de Ciencias, Universidad Central de Venezuela, Caracas 1050, Venezuela and Instituto Universitario de F\'isica
Fundamental y Matem\'aticas, Universidad de Salamanca 37007, Salamanca, Spain}
\author{J. Martin}
\email{chmm@usal.es}
\affiliation{Instituto Universitario de
Fisica Fundamental y Matematicas, Universidad de Salamanca, Salamanca, Spain}

\begin{abstract}
A general procedure to find static and axially symmetric, interior solutions to the Einstein equations is presented. All the so obtained solutions, verify the energy conditions for a wide range of values of the parameters, and match smoothly to some exterior solution of the Weyl family, thereby representing globally regular models describing non spherical sources of gravitational field. In the spherically symmetric limit, all our models converge to the well known  incompressible perfect fluid solution.The key stone of our approach is based on an ansatz allowing to define the interior metric in terms of the exterior metric functions evaluated at the boundary source. Some particular sources are obtained, and the physical variables of the energy-momentum tensor are calculated explicitly, as well as the geometry of the source in terms of the relativistic multipole moments. The total mass of different configurations is also calculated, it is shown to be equal to
 the monopole of the exterior solution.
\end{abstract}
\date{\today}
\maketitle

\section{Introduction}
The search  for exact and physically meaningful analytical solutions to Einstein equations, describing non--spherical sources, is an  endeavour of utmost relevance in general relativity (see \cite{H1} for a discussion on this point).

Indeed, the bifurcation (implied by the Israel theorem \cite{israel}) between the exact spherically symmetric situation (Schwarzschild) and any other exact static solution representing deviations from it, overrides  the fact that deviations from spherical symmetry in compact
self-gravitating objects (white dwarfs, neutron stars), are
likely to be incidental rather than basic features of these systems, and reinforces the need to have available exact non--spherical interiors, in order to study the physical behaviour of very compact objects.

As  it should be expected, this issue has already been considered by many authors in the past. Without  attempting  to be exhaustive, let us  mention the pioneering  work by Hernandez  Jr. \cite{1}, where a general method for obtaining solutions describing axially symmetric sources is presented. Such a method, or some of its modifications  were used in \cite{3} - \cite{7}, to find sources of different Weyl space--times.

This problem has also been considered in  \cite{2}-\cite{r3}. However in all these last references the  line element has been assumed to satisfy the so called Weyl gauge. Now, the fact is that the Weyl gauge is obtained from the condition  $G^\rho_{\rho}+G^z_{z}=0$ (where $G$ denotes the Einstein tensor). Such a condition can always be satisfied in the vacuum (static and axially symmetric)  case. However, for  the interior spacetime, it implies $T^\rho_{\rho}+T^z_{z}=0$, which  represents a restriction on possible solutions.

 More recently, perfect fluid sources have been presented in \cite{que}. However, there exist published results   indicating that static, perfect fluid (isotropic in pressure) sources are spherical (see \cite{npf}  and references therein). This has been confirmed recently for the incompressible isotropic (perfect fluid) in \cite{H1}, where it was shown that such a fluid distribution cannot be matched   to any Weyl exterior, even though it has a surface of vanishing pressure. Accordingly we shall assume here principal stresses to be unequal.

In this work we tackle the problem of finding interior solutions to Einstein equations, by focusing on the matching of the interior space--time to a given exterior (vacuum) solution belonging to the Weyl family, on the boundary surface of the fluid configuration. 

This is a key issue, since only globally  defined solutions (inside and outside the source), may be properly considered as physically meaningful sources of the gravitational field. Otherwise we are bound to introduce shells of matter on the boundary, which would involve physical variables endowed with a dubious interpretation.

As the main ansatz of our approach, we shall write the interior line element in terms of exterior metric functions evaluated on the boundary surface of the source. Doing so, we shall be able to ensure the fulfillment of junction conditions for any of the interiors obtained in this way.  On the other hand, the exterior solutions may be expressed in terms of the relativistic multipole moments (RMM) \cite{geroch}--\cite{th}, which in turn can be measured by means of experiments using gyroscopes or test particles (see for example \cite{herrera}, \cite{h2},  \cite{h3} ). This fact allows us to relate different physical variables describing the source with the multipole moments of the exterior space--time, once we have a globally defined solution.

In order to illustrate our method, we have obtained the sources for two well known static exteriors, belonging to the family of the Weyl metric. The physical, as well as the geometrical properties of these sources are studied in detail, and the range of the values of the parameters, for which these sources describe physically meaningful situations, is established.
\section{The global model of a self-gravitating source}

\subsection{The exterior metric}

The general line element for a vacuum static axially symmetric space--time, in Weyl canonical coordinates may be written as :
\begin{equation}
 ds^2_E=-e^{2\psi}dt^2+e^{-2\psi+2\Gamma}(d\rho^2+dz^2)+e^{-2\psi}\rho^2d\phi^2,
\label{1}
\end{equation}
where $\psi=\psi(\rho,z)$ and  $\Gamma=\Gamma(\rho,z)$ are functions of their arguments.

For vacuum space--times, Einstein's field equations imply for the
metric functions
\begin{equation} \psi_{, \rho \rho}+\rho^{-1}
\psi_{, \rho}+\psi_{, zz} = 0, \label{meq1}
\end{equation}
and
\begin{equation}
\Gamma_{, \rho}= \rho (\psi_{, \rho}^2-\psi_{, z}^2) ; \qquad
\Gamma_{, z}= 2 \rho \psi_{, \rho} \psi_{, z}. \label{meq2}
\end{equation}

Notice  that  (\ref{meq1}) is just the Laplace equation for $\psi$
(in  2--dimensional Euclidean space), and, furthermore, it is precisely
the integrability condition of (\ref{meq2}), that is: given any  $\psi$ satisfying (\ref{meq1}),
a function $\Gamma$ satisfying (\ref{meq2}) always exists. This result may be stated as saying that
for any ``Newtonian'' potential there always exists a specific Weyl
metric, a well known result.

The general solution of the Laplace equation (\ref{meq1}) for the function
$\Psi$, endowed with  an asymptotically flat behaviour, results to be
\begin{equation}
\Psi = \sum_{n=0}^{\infty} \frac{a_n}{R^{n+1}} P_n(\cos \Theta),
\label{psi}
\end{equation}
where $R=(\rho^2+z^2)^{1/2}$, $\cos \Theta= z/R$ are Weyl spherical
coordinates and $P_n(\cos \Theta)$ are Legendre Polynomials. 

The coefficients $a_n$ are arbitrary real constants
 which have been named in the literature ``Weyl moments'', although they
cannot be identified as relativistic multipole moments in spite of the
formal similarity between expression (\ref{psi}) and the Newtonian potential. However these ``Weyl moments''   $a_n$, which provide the so
called ``Newtonian image'' of the solution,  can be  expressed as functions of
the
RMM  \cite{fhp, sueco2, bakdal, sueco3}. Although the full  relations linking
both sets of coefficients
are extremely complicated, they can be used to obtain relatively simple formulas
for the coefficients $\{a_n\}$ in situations where the deviation of the
relativistic solution from spherical symmetry is small. This issue has been
discussed in some detail in \cite{tesis}, \cite{yo}, \cite{sueco}. 

We can write the line element above, in  Erez-Rosen \cite{erroz}, or standard Schwarzschild--tye coordinates $\{r,y\equiv \cos\theta\}$ or in  spheroidal prolate coordinates $\{x\equiv\frac{r-M}{M},y\}$ \cite{quev}:
\begin{equation}
\rho^2=r(r-2M)(1-y^2) \ , \quad z=(r-M)y,
\label{2}
\end{equation}
where $M$ is a constant which will be identified later.

In these prolate spheroidal coordinates, $\Psi$ takes the form
\begin{equation}
\Psi = \sum_{n=0}^{\infty} (-1)^{n+1} q_n Q_n(x) P_n(y),
\label{propsi}
\end{equation}
being $Q_n(y)$ Legendre functions of second kind and $q_n$ a set of
arbitrary constants.
The corresponding expression for the function $\Gamma$, has been obtained
by Quevedo \cite{quev}.

In terms of the above coordinates the line element (\ref{1}) may be writen as:
\begin{widetext}
\begin{eqnarray}
 ds^2_E=-e^{2\psi(r,y)}dt^2+e^{-2\psi+2\left[\Gamma(r,y)-\Gamma^s
 \right]}dr^2+
 e^{-2\left[\psi-\psi^s\right]+
 2\left[\Gamma(r,y)-\Gamma^s\right]}r^2 d\theta^2
+ e^{-2\left[\psi-\psi^s\right]} r^2\sin^2\theta d\phi^2,
\label{exterior}
\end{eqnarray}
\end{widetext}

where  $\psi^s$ and $\Gamma^s$ are the metric functions corresponding to the Schwarzschild solution, namely,
\begin{equation}
\psi^s=\frac 12 \ln \left(\frac{r-2M}{r}\right) \, \quad \Gamma^s=-\frac 12 \ln \left[\frac{(r-M)^2-y^2M^2}{r(r-2M)}\right],
\label{schwfun}
\end{equation}
where the parameter $M$ is easily identified as the Schwarzschild mass.
\subsection{The interior metric}
Let us now consider an interior solution for a spherically symmetric distribution of perfect fluid, which we write generically as:
\begin{equation}
ds^2=-Z(r)^2dt^2+\frac {1}{A(r)} dr^2+r^2\left(d\theta^2+ \sin^2\theta d\phi^2\right),
\label{esferico}
\end{equation}
with $A(r)\equiv 1-pr^2$  and  $Z\equiv\displaystyle{ \frac 32 \sqrt{A(r_{\Sigma})}-\frac 12 \sqrt{ A(r)}}$, and where $p$ is an arbitrary constant and the boundary surface of the source is defined by $r=r_{\Sigma}=const.$ All this corresponds to the well known incompressible (homogeneous energy density) perfect fluid sphere.

The matching of (\ref{esferico}) with the Schwarzschild solution implies $p=\displaystyle{\frac{2M}{r_{\Sigma}^3}}$.

Inspired in the form of (\ref{esferico}), we shall now assume for the interior axially symmetric line element:

\begin{eqnarray}
ds^2_I&=&-e^{2 \hat a} Z(r)^2 dt^2+\frac{e^{2\hat g-2\hat a}}{A(r)} dr^2+e^{2\hat g-2\hat a}r^2 d\theta^2\nonumber \\&+&e^{-2 \hat a}r^2 \sin^2\theta d\phi^2,
\label{interior}
\end{eqnarray}
with
\begin{equation}
\hat a \equiv a(r,\theta)-a^s(r) \ , \qquad  \hat g \equiv g(r,\theta)-g^s(r,\theta),
\label{funcint}
\end{equation}
 where $a^s(r)$ and  $g^s(r,\theta)$ are functions that, on the boundary surface, equal the metric functions corresponding to the Schwarzschild solution  (\ref{schwfun}), i.e. $a^s(r_{\Sigma})=\psi^s_{\Sigma}$ and  $g^s(r_{\Sigma})=\Gamma^s_{\Sigma}$.

It should be noticed  that for  simplicity we consider here only matching surfaces of the form $r = r_\Sigma =
const$, of course more general surfaces with axial symmetry could however be considered as well.

That (\ref{interior}) represents a general (non--spherical) axially symmetric space--time, can be easily seen by transforming such a line element to the general axially symmetric line element, by means of the coordinate transformation $\left\lbrace\displaystyle{ r=\frac{\bar r}{\epsilon \bar r^2+\delta}}\right\rbrace$ , leading to:
\begin{equation}
d\bar s^2_I=-e^{2\nu} dt^2+e^{2 \sigma-2 \nu} \left(d\bar r^2+\bar r^2 d \theta^2\right)+e^{-2\nu} f^2\bar r^2 \sin^2\theta d\phi^2,
\end{equation}
with
\begin{eqnarray}
\nu(\bar r, \theta)\equiv\hat a(\bar r, \theta)+L(\bar r),\\ \nonumber
\sigma(\bar r, \theta)\equiv\hat g(\bar r, \theta)+N(\bar r),\\ \nonumber
f^2\equiv e^{2N(\bar r)} ,
\end{eqnarray}
and 
\begin{equation}
L=\ln\left(\frac{\alpha\bar r^2+\beta}{\epsilon \bar r^2+\delta}\right) \, \quad N=\ln\left[\frac{\alpha\bar r^2+\beta}{(\epsilon \bar r^2+\delta)^2}\right],
\end{equation}
where the parameters $\alpha, \beta,\epsilon, \delta$ have to verify the following constraints : $\alpha \delta-\epsilon \beta= -\delta \epsilon$ 
and  $p\equiv 4 \epsilon \delta$, so that the boundary surface is defined by $r=r_{\Sigma}=\displaystyle{\frac{\sqrt{2\delta\epsilon-\beta\alpha}}{3\delta \epsilon}}$ or $\bar r=\bar r_{\Sigma}=\displaystyle{\sqrt{\frac{\beta+\delta}{\epsilon-\alpha}}}$.

\subsection{The Matching conditions}

We shall now turn to the matching (Darmois) conditions \cite{26}. Thus the continuity of the first and the second fundamental form across the boundary surface implies the continuity of the metric functions and the continuity of the first derivatives  $\partial_r g_{tt}$, $\partial_r g_{\theta \theta}$, $\partial_r g_{\phi \phi}$, producing:
\begin{eqnarray}
&& a_{\Sigma}=\psi_{\Sigma} \ , \quad a^{\prime}_{\Sigma}=\psi^{\prime}_{\Sigma} \ , \quad g_{\Sigma}=\Gamma_{\Sigma} \ , \quad g^{\prime}_{\Sigma}=\Gamma^{\prime}_{\Sigma}, \nonumber\\ 
&&a^s_{\Sigma}=\psi^s_{\Sigma} \ , \quad (a^s)^{\prime}_{\Sigma}=(\psi^s)^{\prime}_{\Sigma}, \nonumber \\ && g^s_{\Sigma}=\Gamma^s_{\Sigma} \ , \quad (g^s)^{\prime}_{\Sigma}=(\Gamma^s)^{\prime}_{\Sigma},
\label{matchingcond}
\end{eqnarray}
where prime denote partial derivative with respect to $r$  and subscript $\Sigma$ indicates that the quantity is evaluated on the boundary surface. It is important to keep in mind that we are using global coordinates $\{r,\theta\}$  on both sides of the boundary.

Thus, our line element (\ref{interior}) matches smoothly with any Weyl exterior  (\ref{exterior}), provided conditions(\ref{matchingcond}) are satisfied.

In the particular case when we want to match our interior with the Schwarzschild exterior, then $\psi=\psi^s$ and $\Gamma=\Gamma^s$, and the source is a perfect fluid  if  $\hat a=\hat g=0$. 
 
However, it is worth noticing  that  the metric functions $\hat a$ and $\hat g$ do not necessarily vanish  in general, as it is apparent from  (\ref{funcint}). Thus in the most general case (within the particular case when the exterior is Schwarzschild) we have a non--spherical source matched with the Schwarzschild space--time.At this point we should recall that  examples of non--spherical sources smoothly matching with the Schwarzschild space--time, are well known in the literature. The most notorious, probably being the matching of the Szekeres metric \cite{Sz1} \cite{Sz2}, as it was shown by Bonnor many years ago \cite{bon2}.

However,  if we demand the source to be represented by a perfect fluid, then we must choose  $\hat a=\mathbb{F}(r)$ and  $\hat g=\mathbb{G}(r)$  since, as already mentioned in the Introduction, static, perfect fluid (isotropic in pressure) sources are spherical (see \cite{npf}. In this latter case we recover the homogeneous density solution  $\mathbb{F}=\mathbb{G}=0$

We shall now see how the field equations constraint further our possible interiors.

\subsection{The field equations and constraints}

Let us first  analyse the well known case when the interior is spherically symmetric, then   $\hat a=\hat g=0$, and the physical variables are obtained from the field equations for a perfect fluid, the result is well known and reads (in relativistic units)
\begin{eqnarray}
-T^0_0\equiv \mu&=&\frac{3p}{8 \pi},\nonumber\\
T^1_1=T^2_2=T^3_3\equiv P&=&\mu\left(\frac{\sqrt A-\sqrt{A_{\Sigma}}}{3 \sqrt{A_{\Sigma}}-\sqrt{A}}\right),
\label{eeesf}
\end{eqnarray}
with  $A=\displaystyle{1-\frac{2m(r)}{r}=1-p r^2=1-\frac{2M r^2}{r_{\Sigma}^3}}$,
where $\mu$ and $P$ denote the energy density and the isotropic pressure respectively, and  for the mass function $m(r)$ we have
\begin{equation}
m(r)=-4\pi\int^{r}_0r^2 T^0_0 dr,
\end{equation}
implying 
\begin{equation}
M\equiv m(r_{\Sigma})=-4\pi\int^{r_{\Sigma}}_0r^2 T^0_0 dr=\frac{p r_{\Sigma}^3}{2}.
\end{equation}

This model, which describes the well known incompressible perfect fluid sphere, is further restricted by the requirement that the pressure be regular and positive everywhere within the fluid distribution, which implies $\displaystyle{\tau\equiv{\frac{r_{\Sigma}}{M}}>\frac94}$. As it is evident from (\ref{eeesf}) the pressure vanishes at the boundary surface. 

Finally, if we impose the strong energy condition $P<\mu$, we should further restrict our model with the condition $\displaystyle{\tau>\frac83}$.

 We shall now proceed to consider the general, non--spherical case. Thus, for our line element  (\ref{interior}) we have the following non vanishing components of the energy--momentum tensor:
\begin{eqnarray}
-T^0_0&=&\kappa \left(8 \pi \mu+\hat p_{zz}-E\right),\nonumber\\
T^1_1&=& \kappa \left(8 \pi P-\hat p_{xx}\right),\nonumber \\
T^2_2&=& \kappa \left(8 \pi P+\hat p_{xx}\right),\nonumber \\
T^3_3&=&\kappa \left(8 \pi P-\hat p_{zz}\right),
\label{eegeneral}
\end{eqnarray}
\begin{widetext}
\begin{equation}
T_1^2=g^{\theta\theta}T_{12}=-\frac{\kappa}{r^2}\left[2 {\hat a}_{,\theta}  \hat a^{\prime}-\hat g^{\prime}\frac{\cos\theta}{\sin\theta}-\frac{\hat g_{,\theta}}{r}+\frac{(1-A)}{r \sqrt A (3 \sqrt{A_{\Sigma}}- \sqrt{A})}(2 {\hat a}_{,\theta} -{\hat g}_{,\theta}  )\right],
\label{pxy}
\end{equation}
\end{widetext}
with $\displaystyle{\kappa\equiv \frac{e^{2\hat a-2\hat g}}{8 \pi}}$, and
\begin{eqnarray}
&&E=-2 \Delta \hat a+(1-A)\left[2 \frac{\hat a^{\prime}}{r}\frac{9 \sqrt{A_{\Sigma}}-4 \sqrt{A}}{3 \sqrt{A_{\Sigma}}- \sqrt{A}}+2 \hat a^{\prime \prime}-\hat g^{\prime \prime}\right],\nonumber\\
&&\Delta \hat a= \hat a^{\prime \prime}+2\frac{\hat a^{\prime}}{r}+\frac{{\hat a}_{,\theta \theta} }{r^2}+\frac{{\hat a}_{,\theta} }{r^2}\frac{\cos \theta}{\sin \theta},\nonumber \\
&&\hat p_{xx}=-\frac{{\hat a}_{,\theta} ^2}{r^2}-\frac{\hat g^{\prime}}{r}+\hat a^{\prime 2}+\frac{{\hat g}_{,\theta} }{r^2}\frac{\cos \theta}{\sin \theta}+\nonumber\\
&+&(1-A)\left[2 \frac{\hat a^{\prime}}{r}\frac{\sqrt{A}}{3 \sqrt{A_{\Sigma}}- \sqrt{A}}- \hat a^{\prime 2} +\frac{\hat g^{\prime}}{r}\frac{3 \sqrt{A_{\Sigma}}-2 \sqrt{A}}{3 \sqrt{A_{\Sigma}}-\sqrt A}\right], \nonumber \\
&&\hat p_{zz}=-\frac{{\hat a}^2_{,\theta} }{r^2}-\frac{\hat g^{\prime}}{r}-\hat a^{\prime 2}-\frac{{\hat g}_{,\theta \theta} }{r^2}-\hat g^{\prime \prime}+\nonumber\\
&+&(1-A)\left[-2 \frac{\hat a^{\prime}}{r}\frac{\sqrt{A}}{3 \sqrt{A_{\Sigma}}- \sqrt{A}}+ \hat a^{\prime 2} +2\frac{\hat g^{\prime}}{r}\right],
\label{eegeneraldet}
\end{eqnarray}
where $\hat p_{zz}, \hat p_{xx}, T_1^2, E$ describe deviations from the spherical symmetry.
Indeed, if $\hat a=\hat g=0$,  then $E=\hat p_{xx}=\hat p_{zz}=0$ and we recover the  spherical case.
 
From the expressions above, using   (\ref{eegeneral})-(\ref{eegeneraldet}) and  introducing the dimensionless parameter $s\equiv r/r_{\Sigma}$, we can now obtain the explicit expressions for the physical variables:
\begin{widetext}
\begin{eqnarray}
-T^0_0=\frac{\kappa}{r_{\Sigma}^2}\left\lbrace\frac{6}{\tau}-\frac{{\hat a_{,\theta}}^2}{s^2}-A\left[ (\partial_s\hat a)^2 +(\partial_s^2\hat g)-2(\partial_s^2\hat a)\right]
-\frac{{\hat g_{,\theta \theta}}}{s^2}+(1-2A)\frac{\partial_s \hat g}{s}
-  2(1-3A) \frac{\partial_s \hat a}{s}+\frac{2}{s^2}\left({\hat a_{,\theta \theta}}+{\hat a_{,\theta}}\frac{\cos\theta}{\sin\theta} \right)
\right\rbrace,
\label{tdens}
\end{eqnarray}
\end{widetext}

\begin{widetext}
\begin{eqnarray}
T_1^1=\frac{\kappa}{r_{\Sigma}^2}\left\lbrace\frac{6}{\tau}\frac{\sqrt A-\sqrt{A_{\Sigma}}}{3\sqrt{A_{\Sigma}}-\sqrt A}+\frac{{\hat a_{,\theta}}^2}{s^2}-(\partial_s\hat a)^2 A-\frac{{\hat g_{,\theta}}}{s^2}\frac{\cos\theta}{\sin\theta}- 2 \frac{\partial_s \hat a}{s}\frac{\sqrt A (1-A)}{3\sqrt{A_{\Sigma}}-\sqrt A}-\frac{\partial_s \hat g}{s}\left[\frac{(1-A)(3\sqrt{A_{\Sigma}}-2\sqrt A)}{3\sqrt{A_{\Sigma}}-\sqrt A}-1 \right]\right\rbrace,
\label{t11}
\end{eqnarray}
\end{widetext}

\begin{widetext}
\begin{eqnarray}
T_3^3=\frac{\kappa}{r_{\Sigma}^2}\left\lbrace\frac{6}{\tau}\frac{\sqrt A-\sqrt{A_{\Sigma}}}{3\sqrt{A_{\Sigma}}-\sqrt A}+\frac{{\hat a_{,\theta}}^2}{s^2}+(\partial_s\hat a)^2 A+(\partial^2_{ss}\hat g)+\frac{{\hat g_{,\theta \theta}}}{s^2}+ 2 \frac{\partial_s \hat a}{s}\frac{\sqrt A (1-A)}{3\sqrt{A_{\Sigma}}-\sqrt A}-\frac{\partial_s \hat g}{s} (1-2A) \right\rbrace,
\label{t33}
\end{eqnarray}
\end{widetext}

\begin{widetext}
\begin{equation}
T_1^2=-\frac{\kappa}{s^2r_{\Sigma}^3}\left[2 {\hat a}_{,\theta}  \partial_s\hat a-\partial_s\hat g\frac{\cos\theta}{\sin\theta}-\frac{\hat g_{,\theta}}{s}+\frac{2s (2 {\hat a}_{,\theta} -{\hat g}_{,\theta}  )}{ \sqrt{\tau-2s^2} (3 \sqrt{\tau-2}- \sqrt{\tau-2s^2})}\right].
\label{t12}
\end{equation}
\end{widetext}
 Obviously for any specific (non--spherical) model we need to provide explicit forms for  $\hat a$ and $\hat g$, however even at this level of generality we can assure that the junction conditions  (\ref{matchingcond}) imply $(P_{rr}\equiv g_{rr}T^1_1)_{\Sigma}=0$. 

We shall first proceed to proof the above statement, and then we shall provide a general procedure to choose $\hat a$ and $\hat g$ producing physically meaningful  models.

It is always possible to choose the metric functions $\hat a$ and  $\hat g$ such that, once the junction conditions (\ref{matchingcond}) are satisfied, the angular derivatives of such functions are continuous, i.e.:  $({\hat a}_{,\theta})_{\Sigma}=({\psi}_{,\theta})_{\Sigma}$ and  $({\hat g}_{,\theta})_{\Sigma}=({\Gamma}_{,\theta})_{\Sigma}$.

Then, using $A_{\Sigma}=\displaystyle{\frac{r_{\Sigma}-2M}{r_{\Sigma}}}$, we obtain for  $\hat p_{xx}$  on the boundary surface:
\begin{widetext}
\begin{equation}
(\hat p_{xx})_{\Sigma}=\frac{1}{r_{\Sigma}^2} \left[  -({\hat{\psi}}_{,\theta})_{\Sigma}^2 +({\hat{\Gamma}}_{,\theta})_{\Sigma} \frac{\cos\theta}{\sin\theta}+2M\hat{\psi}^{\prime}_{\Sigma}+
r_{\Sigma}(r_{\Sigma}-2M)\hat{\psi}^{\prime 2}_{\Sigma}\ -(r_{\Sigma}-M)\hat{\Gamma}^{\prime}_{\Sigma}\right],
\label{pxxe}
\end{equation}
\end{widetext}
where $\hat{\psi}_{\Sigma}\equiv \psi_{\Sigma}-\psi^s_{\Sigma}$ and  $\hat{\Gamma}_{\Sigma}\equiv \Gamma_{\Sigma}-\Gamma^s_{\Sigma}$.

Taking into account   the Einstein's equations, we find for $\Gamma^{\prime}$ and $\Gamma_{,\theta}$  the following expressions
\begin{widetext}
\begin{eqnarray}
\Gamma^{\prime}&=& \frac{\sin^2\theta}{(r-M)^2-M^2\cos^2\theta}\left[r(r-M)(r-2M)(\psi^{\prime})^2-(r-M){ \psi^2_{,\theta}}+2\frac{\cos\theta}{\sin\theta} r (r-2M) \psi^{\prime} { \psi_{,\theta}} \right],\nonumber \\
\Gamma_{,\theta}&=&\frac{-\sin\theta r(r-2M)}{(r-M)^2-M^2\cos^2\theta}\left[r(r-2M)\cos\theta (\psi^{\prime})^2-\cos\theta{ \psi^2_{,\theta}}-2\sin\theta  (r-M) \psi^{\prime} { \psi_{,\theta}} \right],
\end{eqnarray}
\end{widetext}
then producing the vanishing of $(\hat p_{xx})_{\Sigma}$.

In a similar way it can be shown that  $T_1^2$ vanishes on the boundary surface. This last condition, which follows from the Darmois conditions,  and therefore  is necessary, in order to avoid the presence of shells on the boundary surface, can be obtained  at once from a simple inspection of the equation (22) in \cite{H1}.

We shall now  calculate  the total mass for any of our models. We shall see that it  always coincides with the relativistic monopole  of the exterior solution. Therefore, whenever $q_0=1$, as in the Erez--Rosen or the $M-Q$ metric (see the next section), it will coincide with the mass parameter of the Schwarzschild metric. However for the $\gamma$ metric, $q\neq 1$, and the total mass does not coincide with the  Schwarzschild mass (see the next section).

 As  is well known, both the Komar mass $M_K$  and the  Tolman $M_T$ coincide for the static case. 

Thus

\begin{equation}
M_T=M_K=-\frac{1}{4\pi}\int_V\sqrt{-\bf{g}}R^0_0 d^3x=\int_V(-T^0_0+\hat T)\sqrt{-\bf{g}} d^3x,
\end{equation}
with an obvious notation.

Taking into account 
\begin{eqnarray}
\sqrt{-\bf{g}}R^0_0&=&\partial_k\left(\sqrt{-\bf{g}}g^{00}\Gamma^k_{00}\right)=
-\partial_k \left(-\sqrt{\bf{\hat g}}\hat g^{kj}\partial_j\sqrt{-g_{00}}\right)\nonumber \\&=&-\sqrt{\bf{\hat g}}\hat{\Delta} \sqrt{-g_{00}},
\end{eqnarray}
where  $\bf{\hat{g}}$ is the determinant of the three--space metric and  $\hat{\Delta}$ denotes the second kind  Beltrami operator for such a metric.

Then we may write for the  total mass
\begin{equation}
M_T=M_K=\int_V \rho_T \hat{\eta},
\end{equation}
where  $\hat{\eta}=\sqrt{\bf{\hat g}}d^3x$ is the three dimensional volume element, and  
\begin{equation}
\hat{\Delta}\sqrt{-g_{00}}=4\pi \rho_T \ , \quad \rho_T\equiv\sqrt{-g_{00}} \left(-T^0_0+\hat T\right).
\end{equation}

Next, applying the  Gauss theorem
\begin{equation}
M_T=\frac{1}{4\pi}\int_V\hat{\Delta}\sqrt{-g_{00}}\hat{\eta}=\frac{1}{4\pi}\int_V\partial_k\left(-\sqrt{\bf{\hat g}}\hat g^{kj}\partial_j\sqrt{-g_{00}}\right) d^3x,
\end{equation}
we obtain for the total mass
\begin{equation}
M_T=\frac{1}{4\pi}\int_{\partial V}\sqrt{\bf{\hat g}} \hat g^{kj}\partial_j\sqrt{-g_{00}} \ n_k d\sigma.
\end{equation}

The above expression yields, for our interior (\ref{interior}) 
\begin{equation}
M_T=\frac 12 \int_0^{\pi} r_{\Sigma}^2 \sqrt{A_{\Sigma}}e^{-\hat{\psi}_{\Sigma}}
 \left(\partial_r\sqrt{-g_{00}}\right)_{\Sigma} \sin\theta d \theta ,
\label{mn1}
\end{equation}

In the particular case when we match our source with the  Schwarzschild exterior, we have  $\hat{\psi}_{\Sigma}=0$, and  $\sqrt{-g_{00}}=Z(r)$, implying  $\displaystyle{\left(\partial_r\sqrt{-g_{00}}\right)_{\Sigma}=\frac{M}{r_{\Sigma}^2\sqrt{A_{\Sigma}}}}$ and accordingly $M_T=M$.

In the general (non--spherical) case,  $\hat{\psi}_{\Sigma}=\psi_{\Sigma}-\frac 12 \ln A_{\Sigma}$ where $\psi_{\Sigma}$ is any exterior metric function belonging to the Weyl family, and $\sqrt{-g_{00}}=e^{\hat a}Z(r)$, which implies 
\begin{equation}
\left(\partial_r \sqrt{-g_{00}}\right)_{\Sigma}=e^{\hat{\psi}_{\Sigma}}\left(
\frac{M}{r_{\Sigma}^2\sqrt{A_{\Sigma}}}+\sqrt{A_{\Sigma}} \hat{\psi}_{\Sigma}^{\prime}\right),
\end{equation}
then, feeding back the above expression into (\ref{mn1}) we obtain
\begin{equation}
M_T=M+\frac{1}{2}r_{\Sigma}^2 A_{\Sigma}\int_0^{\pi}d\theta \sin\theta \hat{\psi}_{\Sigma}^{\prime}.
\label{masatotal}
\end{equation}

Now, we know that all the solutions belonging to the Weyl family, may be written as (\ref{propsi}), where the first term in the series, corresponds to the Schwarzschild  solution $\psi^s$. 

Then, we may write

\begin{equation}
\hat{\psi}^{\prime}_{\Sigma}=
\psi^{\prime}_{\Sigma}-(\psi^s)^{\prime}_{\Sigma}=-\sum_{n=1}^{\infty}q_nQ_n^{\prime}(-1+r/M)_ {\Sigma}P_n(y),
\label{m1}
\end{equation}
and taking into account the orthogonality relation of the Legendre  polynomial 
\begin{equation}
\int_{-1}^1dy P_n(y)=2 \delta_{0n} \ , \forall  n,
\label{nm3}
\end{equation}
we obtain that the integral in  (\ref{masatotal}) vanishes identically whenever the exterior solution satisfies the condition $q_0=1$.

\subsection{The ansatz for the metric functions}

We shall now get back to the problem of choosing the functions  $\hat a$ and  $\hat g$, leading to physically meaningful solutions. With this aim,  besides the fulfillment of the junction conditions (\ref{matchingcond}), we shall require that all physical variables be regular within the fluid distribution and the energy density to be positive.

To ensure the fulfillment of the junction conditions  (\ref{matchingcond}), we may write without loos of generality, 
\begin{eqnarray}
\hat a&=&\hat \psi_{\Sigma}(\theta)+\hat \psi^{\prime}_{\Sigma}(\theta) (r-r_{\Sigma})+\Lambda(r,\theta)(r-r_{\Sigma})^2\nonumber\\
\hat g&=&\hat \Gamma_{\Sigma}(\theta)+\hat \Gamma^{\prime}_{\Sigma}(\theta) (r-r_{\Sigma})+\Xi(r,\theta)(r-r_{\Sigma})^2,
\label{ayg}
\end{eqnarray}
where $ \Lambda(r,\theta)$ and $\Xi(r,\theta)$ are so far two arbitrary functions of their arguments.

On the other hand, to guarantee a good behaviour of the physical variables at the center of the distribution we shall demand: 
\begin{eqnarray}
&\hat a^{\prime}_0={\hat a}_{,\theta 0}={\hat a }_{,\theta  \theta 0}={\hat a}^{\prime}_{,\theta 0}={\hat a }^{\prime}_{,\theta \theta 0}=0 \nonumber \\
&\hat g^{\prime}_0={\hat g}_{,\theta 0}={\hat g }_{,\theta \theta 0}={\hat g}^{\prime}_{,\theta 0}={\hat g }^{\prime}_{,\theta \theta 0}=0 \nonumber\\
&\hat g^{\prime \prime}_0={\hat g}^{\prime \prime}_{,\theta 0}=0,
\label{condcero}
\end{eqnarray}
where (\ref{pxy}, \ref{eegeneraldet}) have been used, and the subscript $0$ indicates that the quantity is evaluated at the center of the distribution. 

Using the conditions above in  (\ref{ayg}) we may write for  $\Lambda$ and  $\Xi$ 
\begin{eqnarray}
\Lambda(r,\theta)&=&\Lambda_0(\theta)+\Lambda^{\prime}_0(\theta) r+F(r,\theta)\nonumber \\
\Xi(r,\theta)&=&\Xi_0(\theta)+\Xi^{\prime}_0(\theta) r+\Xi^{\prime \prime}_0(\theta) r^2 +G(r,\theta)
\end{eqnarray}
with  $F(0,\theta)=F^{\prime}(0,\theta)=0$ and  $G(0,\theta)=G^{\prime}(0,\theta)=G^{\prime \prime}(0,\theta)=0$.
Then we can finally write for $\hat a$ and $\hat g$
\begin{eqnarray}
\hat a(r,\theta)&=&r^2\left(-\frac{\hat \psi^{\prime}_{\Sigma}}{r_{\Sigma}}+3\frac{\hat \psi_{\Sigma}}{r_{\Sigma}^2}\right)+r^3 \left(-\frac{\hat \psi^{\prime}_{\Sigma}}{r_{\Sigma}^2}-2\frac{\hat \psi_{\Sigma}}{r_{\Sigma}^3}\right)\nonumber \\&+&(r-r_{\Sigma})^2F(r,\theta)
\label{aygdefa}
\end{eqnarray}

\begin{eqnarray}
\hat g(r,\theta)&=&r^4\left(\frac{\hat \Gamma^{\prime}_{\Sigma}}{r_{\Sigma}^2}-3\frac{\hat \psi_{\Sigma}}{r_{\Sigma}^3}\right)+r^3 \left(-\frac{\hat \Gamma^{\prime}_{\Sigma}}{r_{\Sigma}^2}+4\frac{\hat \psi_{\Sigma}}{r_{\Sigma}^3}\right)\nonumber \\&+&(r-r_{\Sigma})^2G(r,\theta).
\label{aygdef}
\end{eqnarray}

The so obtained metric functions, satisfy the junction conditions  (\ref{matchingcond}) and produce physical variables which are regular within the fluid distribution.  Furthermore  the vanishing of  $\hat g$ on the axis of symmetry, as required by the regularity conditions,  necessary to ensure elementary flatness in the vicinity of  the axis of symmetry, and in particular at the center (see \cite{1n}, \cite{2n}, \cite{3n}),
 is assured by the fact that $\hat \Gamma_{\Sigma}$ and $\hat \Gamma^{\prime}_{\Sigma}$ vanish on the axis of symmetry. 
So far we have presented the general procedure to build up sources for the Weyl metric, in what follows, we shall illustrate the method with some examples.

\section{Particular solutions}

\subsection{``Schwarzschild interiors''}

We shall first analyse the possible cases of interiors matching with the Schwarzschild exterior on the boundary surface.

From (\ref{aygdefa}) and (\ref{aygdef}) we see that the most general form of the metric functions $\hat a(r,\theta)$ and $\hat g(r,\theta)$ is:

\begin{eqnarray}
&&\hat a(r,\theta)=(r-r_{\Sigma})^2F(r,\theta)\equiv \mathbb{F},\nonumber \\&&\hat g(r,\theta)=(r-r_{\Sigma})^2G(r,\theta)\equiv \mathbb{G}.
\end{eqnarray}
 For the particular case when  $F=G=0$ we recover the perfect fluid case with homogeneous energy density, described previously.  

Therefore, if we look for other perfect fluid solutions  (i.e., $\hat p_{xx}=\hat p_{zz}=T_1^2=0$), then we should assume that the functions $F$ and  $G$  only depend on $r$ (since all static perfect fluids solutions should be spherically symmetric), in which  case the Einstein equations reduce to:
\begin{equation}
\mathbb{G}^{\prime}=0 \ , \qquad  \mathbb{F}^{\prime 2}+(1-A)\left(-\mathbb{F}^{\prime 2}+\frac 2r \mathbb{F}^{\prime} \frac{\sqrt A}{3 \sqrt{A_{\Sigma}}-\sqrt A}\right)=0.
\label{otrofluper}
\end{equation}

The only possible solution to the above system, satisfying the junction conditions  $\mathbb{G}_{\Sigma}=\mathbb{G}^{\prime}_{\Sigma}=0$ and  $\mathbb{F}_{\Sigma}=\mathbb{F}^{\prime}_{\Sigma}=0$,
is  $\mathbb{G}=\mathbb{F}=0$.

Indeed, the second equation in   (\ref{otrofluper}) has only two possible solutions, namely:
\begin{equation}
\mathbb{F}^{\prime}=0 \ , \qquad \mathbb{F}^{\prime}=-2\frac{Z(r)^{\prime}}{Z(r)},
\end{equation}
(where $Z(r)$  is defined in  (\ref{esferico})). The first of which implies   $\mathbb{F}=0$, whereas the second one is incompatible with the junction conditions.

Thus, within the context of our approach, the only possible interior consisting in a perfect fluid and  matching  with the Schwarzschild space--time  on the boundary surface is the spherically symmetric, fluid distribution with  homogeneous energy density. 


The question about the possible existence of  non--spherical sources described by non--perfect fluids and matching with Schwarzschild space--time  on the boundary surface, remains open. In this latter case, both  $\mathbb{F}$ and  $\mathbb{G}$ would depend on $\theta$  and therefore should not be spherically symmetric. However we shall not follow here, this line of research.

Instead, we shall looking for sources matching with exterior solutions of the Weyl family, other than the Schwarzschild space--time.

In what follows we shall construct some interiors which match with some specific known solutions of the Weyl family with the additional assumption $\mathbb{F}=\mathbb{G}=0$ .
\vskip 5mm
\subsection{A source for the exterior $M-Q^{(1)}$ solution}

In \cite{yo} it was shown that it is possible to find a solution of the
Weyl family, by a convenient choice of coefficients
$a_n$, such that the resulting solution possesses only monopole and
quadrupole moments (in the Geroch sense). The obtained solution
($M-Q$)
may be written, as follows:
\begin{equation}
\Psi_{M-Q}=\Psi_{q^0}+q \Psi_{q^1}+q^2 \Psi_{q^2}+\ldots =
\sum_{\alpha=0}^\infty q^\alpha\Psi_{q^\alpha} \quad ,
\label{(15)}
\end{equation}
where the zero-th order corresponds to the
Schwarzschild solution.

It appears that each power in $q$ adds a quadrupole correction to the
spherically symmetric solution. Now, it should be observed that due to the
linearity of Laplace equation,
these corrections give rise to a series of exact solutions. In other
words, the power series of $q$ may be cut at any order, and the partial
summation, up to that order, gives an exact
solution representing a quadrupole correction to the  Schwarzschild solution.

Since we are mainly interested in slight deviations from spherical symmetry, we shall consider the M--Q solution, only up to first order in $q$, hereafter referred as  $M-Q^{(1)}$.

So, let us consider the line element  (\ref{interior}) with   metric functions given by (\ref{aygdefa}) and (\ref{aygdef})
\begin{eqnarray}
\hat a(r,\theta)&=&\hat \psi_{\Sigma} s^2(3-2s)   +r_{\Sigma}\hat \psi^{\prime}_{\Sigma}s^2(s-1),\nonumber \\
\hat g(r,\theta)&=&\hat \Gamma_{\Sigma} s^3(4-3s)   +r_{\Sigma}\hat \Gamma^{\prime}_{\Sigma}s^3(s-1).
\label{aygsimple}
\end{eqnarray}
with $s\equiv r/r_{\Sigma} \in \left[0,1\right]$.

Then, for the  M-Q$^{(1)}$  solution we may write:
\begin{eqnarray}
\hat \psi_{\Sigma}\equiv \psi_{\Sigma}-\psi^s_{\Sigma}=q \psi_{q},\nonumber\\
\hat \Gamma_{\Sigma}\equiv \Gamma_{\Sigma}-\Gamma^s_{\Sigma}=q \Gamma_{q}+q^2\Gamma_{q^2}.
\label{prepsimq1}
\end{eqnarray}
with 
\begin{eqnarray}
\psi_q&=&\frac{5}{32}\left\lbrace\left[(3y^2-1)(3\tau^2-6\tau+2)-4\right]\ln\left(1-\frac{2}{\tau}\right)\right.+\nonumber\\
&-&\left.\frac{8(\tau-1)}{(\tau-1)^2-y^2}+6(\tau-1)(3y^2-1)\right\rbrace,
\label{psimq1}
\end{eqnarray}
\begin{widetext}
\begin{eqnarray}
\Gamma_q=-\frac 58 (1-y^2)\left\lbrace 6+ 3(\tau-1)\ln\left(1-\frac{2}{\tau}\right)+2\frac{(\tau-1)^2+y^2}{[(\tau-1)^2-y^2]^2}\right\rbrace,
\end{eqnarray}
\end{widetext}
\begin{widetext}
\begin{eqnarray}
\Gamma_{q^2}&=&\frac{225}{48}\ln\left[\frac{\tau(\tau-2)}{(\tau-1)^2-y^2}\right]-\frac{75}{16}(1-y^2)\frac{(\tau-1)^2}{(\tau-1)^2-y^2}
+\frac{225}{1024}\ln\left(1-\frac{2}{\tau}\right) (1-y^2)\times\nonumber\\
&\times&\left\lbrace 4(\tau-1)\left[\tau^2-2\tau+2+y^2(-9\tau^2+18\tau-2)-\frac83\frac{\tau^2-2\tau+2}{(\tau-1)^2-y^2}\right]+\right.\nonumber\\
&+&\left.\tau(\tau-2)\left[\tau^2-2\tau+y^2(-8-9\tau^2+18 \tau) \right]\ln\left(1-\frac{2}{\tau}\right)\right\rbrace+\nonumber\\
&+&\frac{225}{256}(1-y^2)\left[\tau^2-2\tau+5+y^2(-5-9\tau^2+18\tau)\right]
-\frac{75}{192}\frac{(1-y^2)}{[(\tau-1)^2-y^2]^4}\times\nonumber\\
&\times&\left[(\tau-1)^4\left( 2\tau^2-4\tau+1+3y^2\right)-y^4-y^6+(\tau-1)^2y^2(-6+4y^2)\right].\nonumber\\
\end{eqnarray}
\end{widetext}

A straightforward calculation, using (\ref{tdens})--(\ref{t12}) allows us to find the explicit expressions for the physical variables, these are displayed in figures (3)--(7). However, before entering into a detailed discussion of these figures, 
we shall  carry out some calculations with the purpose of providing some information about the ``shape'' of the source. In particular we shall see how it is related with the relativistic  quadrupole moment of the source. 

For doing so, using our interior metric, we shall calculate the proper length of the object  along the axis  $l_z$ and the proper equatorial radius $l_{\rho}$:
\begin{equation}
l_z\equiv\int_0^{r_{\Sigma}}\frac{e^{\hat g(y=1)-\hat a(y=1)}}{\sqrt A} dz \ , \  l_{\rho}\equiv\int_0^{r_{\Sigma}}\frac{e^{\hat g(y=0)-\hat a(y=0)}}{\sqrt A} d\rho,
\end{equation}
where ${\rho,z}$ are the cylindrical coordinates associated to the  Erez-Rosen coordinates.

Obviously in the spherical case ($\hat a=\hat g=0$), both lengths are identical:
\begin{equation}
l_z^s=l_{\rho}^s=\int_0^{r_{\Sigma}}\frac{d\xi}{\sqrt{1-p\xi^2}} =r_{\Sigma}\sqrt{\frac{\tau}{2}} \arcsin\sqrt{\frac{2}{\tau}},
\label{lzrho}
\end{equation}
where the fact that  $p=\displaystyle{\frac{2}{\tau r_{\Sigma}^2}}$ has been taken into account and where  $l_z^s$, $l_{\rho}^s$ denote the lengths corresponding to the spherical case. 

 With $l_z$ and  $l_{\rho}$ we may define the  ellipticity  as $e\equiv 1-\frac{l_{\rho}}{l_z}$. The two extreme values of this parameter are $e=0$, which corresponds to a spherical object, and   $e=1$ for the limiting  case when  the source is represented by a disk. In between of these two extremes we have  $e>0$ for a  prolate source and  $e<0$ for an oblate one.

In the general (non--spherical case) we must compare function $\displaystyle{e^{-\hat a(y=1)}}$ with  $\displaystyle{e^{\hat g(y=0)-\hat a(y=0)}}$, since $\hat g(y=\pm 1)$ vanishes along the axis.

 It can be seen that the sign of both  functions is positive, and their relative  magnitudes are  determined by the sign of $q$, no matter the sign of the quadrupole parameter   $q$,  leading to  the well  known result that  $q<0$ implies that the object is oblate (since $l_z<l_{\rho}$), while  $q>0$ implies that it is prolate (since $l_z>l_{\rho}$). In the figure 1, one example is shown.

\begin{figure}[h]
$$
\begin{array}{cc}
 \includegraphics[scale=0.21]{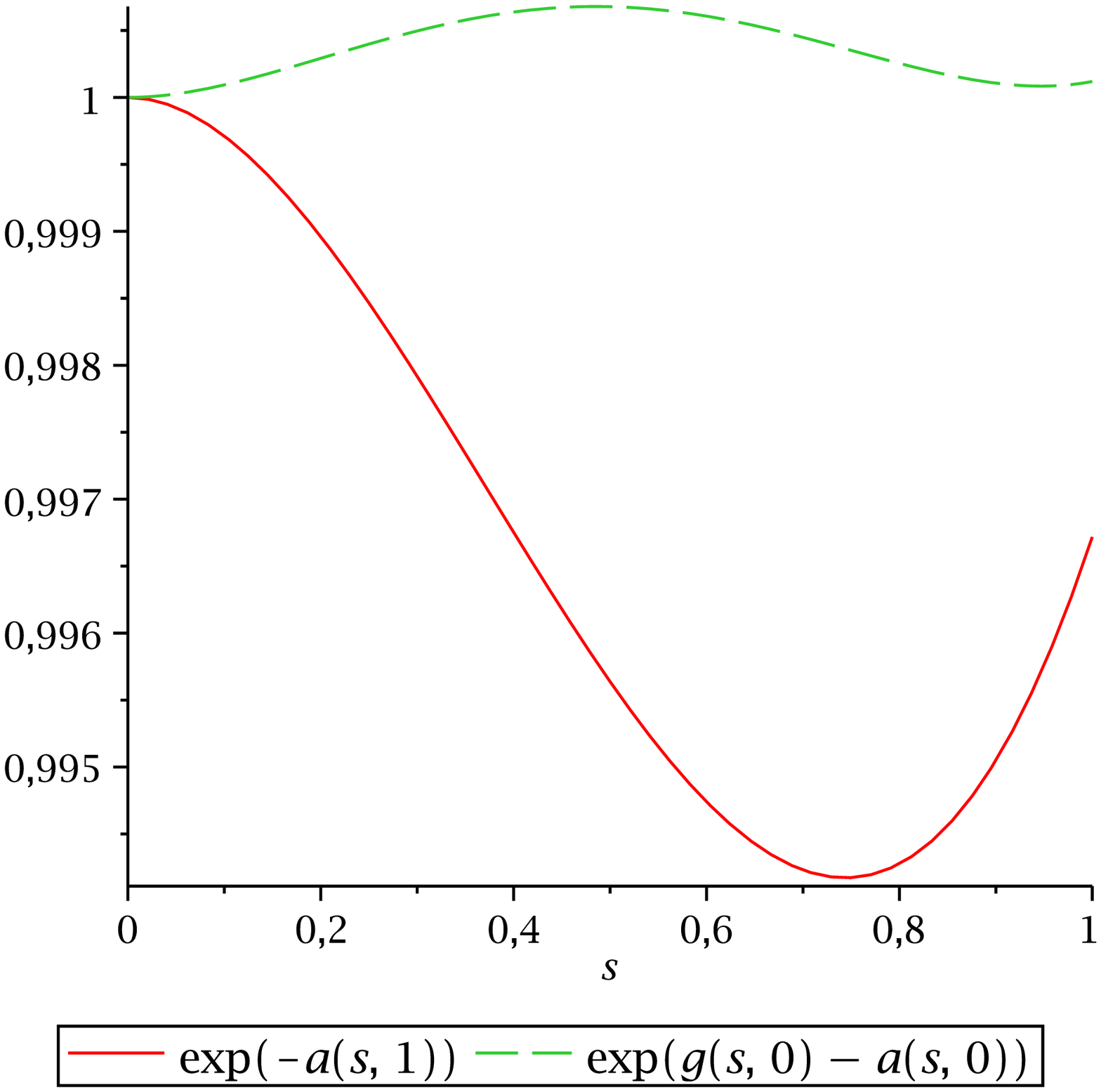}& \includegraphics[scale=0.21]{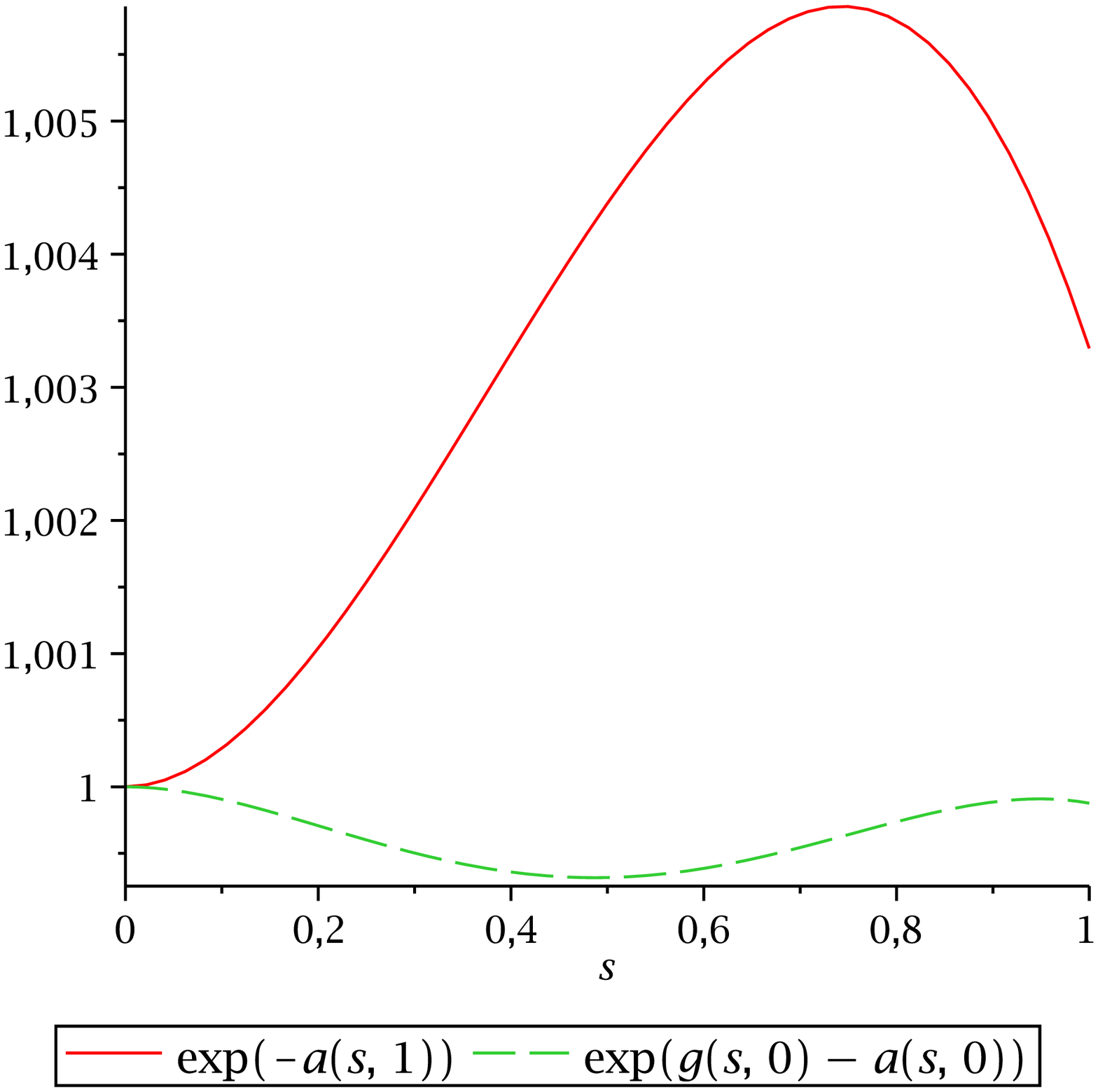}  \nonumber\\
(a) & (b) \nonumber
\end{array}
$$
\caption{\label{longitudes} {\it  Functions  $exp(\hat g(s,y=0)-\hat a(s,y=0))$ and  $exp(-\hat a(s,y=1))$  for positive (curve b, prolate source) and negative (curve a, oblate source) values of the quadrupole parameter $q=\pm 0.01$ and   $\tau=2.7$.}}
\end{figure}

\vskip 5mm

Figure 2 shows the ellipticity $e$  of the source as a function  of the quadrupole parameter $q$, for different values of the  parameter $\tau$. As can be seen, the relation between $e$ and $q$ is increasingly linear for each $\tau$, i.e., the higher is $q$ (positive values) the higher is $e$ and therefore the shape of the source is more prolate (elongated along the axis). Of course, for $q=0$ we recover the sphericity ($e=0$). It is also observed from the figure 2 that the deformation of the source with respect to the spherical case, for any value of $q$, is smaller for larger values of  $\tau$ (less compact object), i.e., the slope of the curves relating $e$ and $q$ decrease with $\tau$.

\begin{figure}[h]
 \includegraphics[scale=0.31]{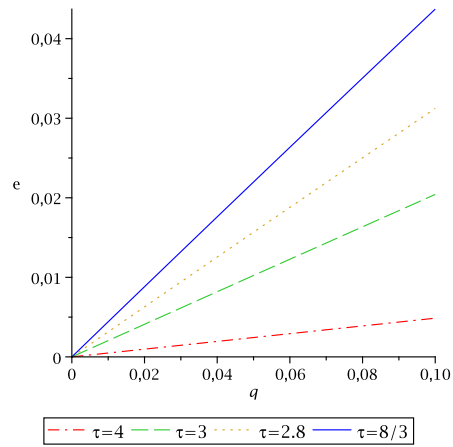}  
\caption{\label{ellip} {\it  Relation between the ellipticity of the source $e$ and its  quadrupole moment $q$ for different values of $\tau$.}}
\end{figure}

We can also calculate the total area of the boundary surface  $S_{\Sigma}$, we obtain
\begin{equation}
S_{\Sigma}=\int_0^{2\pi}d\phi\int_0^{\pi} d\theta \sin\theta r_{\Sigma}^2 e^{\hat g_{\Sigma}-2\hat a_{\Sigma}}
\end{equation}

In the spherically symmetric case  we re-obtain the well know result, $S_{\Sigma}=4\pi r_{\Sigma}^2$, whereas for the general case we get
\begin{equation}
S_{\Sigma}=2\pi r_{\Sigma}^2\int_{-1}^{1} dy  \  e^{\hat \Gamma_{\Sigma}-2\hat \psi_{\Sigma}}.
\label{area}
\end{equation}
Therefore if,  $\hat \Gamma_{\Sigma}-2\hat \psi_{\Sigma}>0$ then  $S_{\Sigma}>4\pi r_{\Sigma}^2$.  As it can be very easily verified, our source for the $M-Q^{(1)}$ solution satisfies the above condition ($S_{\Sigma}>4\pi r_{\Sigma}^2$) for all values, both, positive and negative, of the quadrupole parameter $q$, and any value of $\tau$.

Let us now turn back to the physical variables of our model. Figure  3 exhibits the behaviour of the radial pressure $P_{rr}\equiv g_{rr} T_1^1$  for different values of  $q$ (positive and negative).

In it, we observe the variation of the radial pressure with respect to the spherically symmetric case  ($q=0$). This variation is smaller for angle values close to the equator, as it is apparent for $y=0.3$. Notice that the radial pressure is positive, with negative pressure gradient, and vanishes on the boundary surface.

\begin{figure}[h]
$$
\begin{array}{cc}
 \includegraphics[scale=0.21]{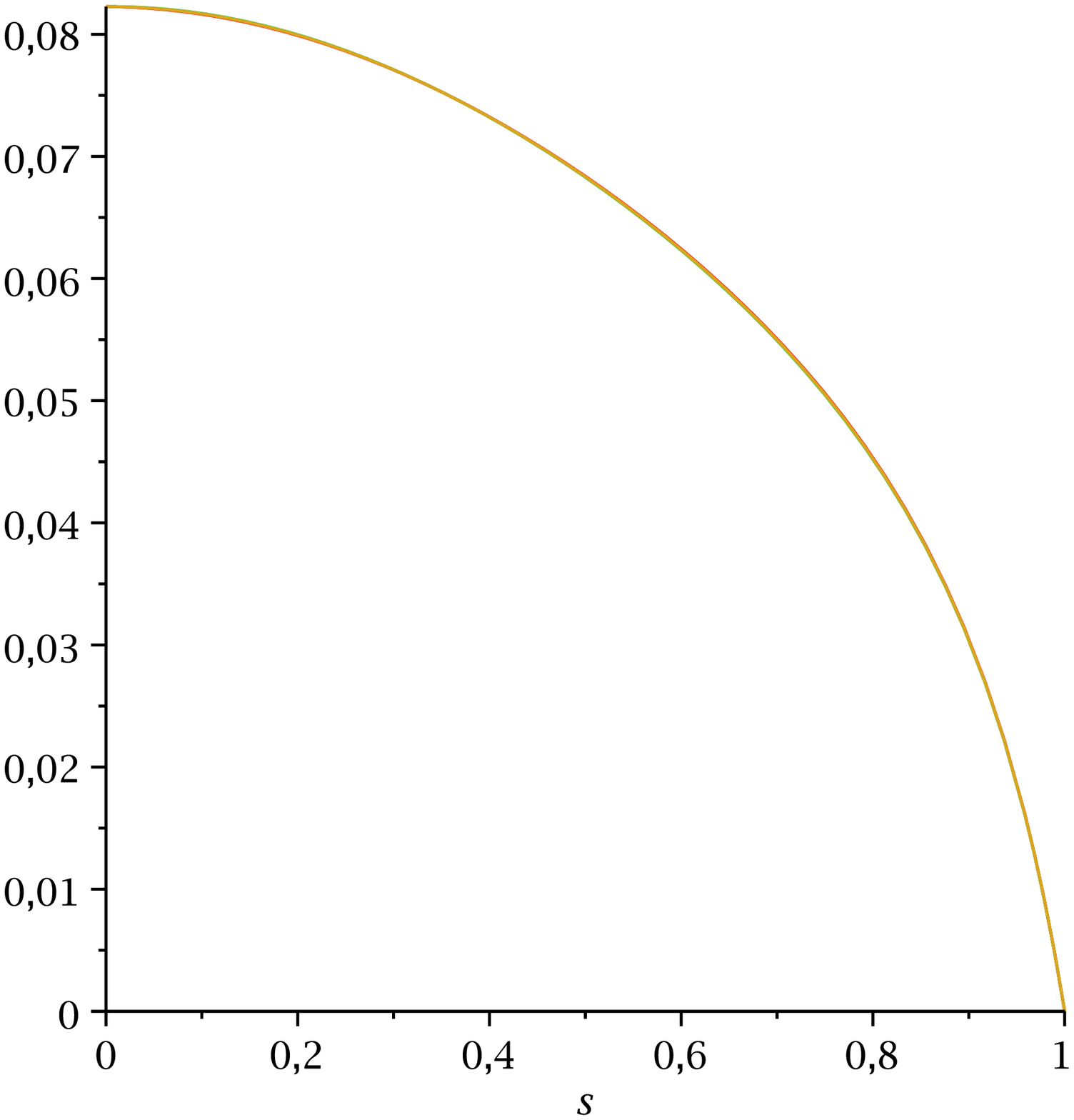}& \includegraphics[scale=0.21]{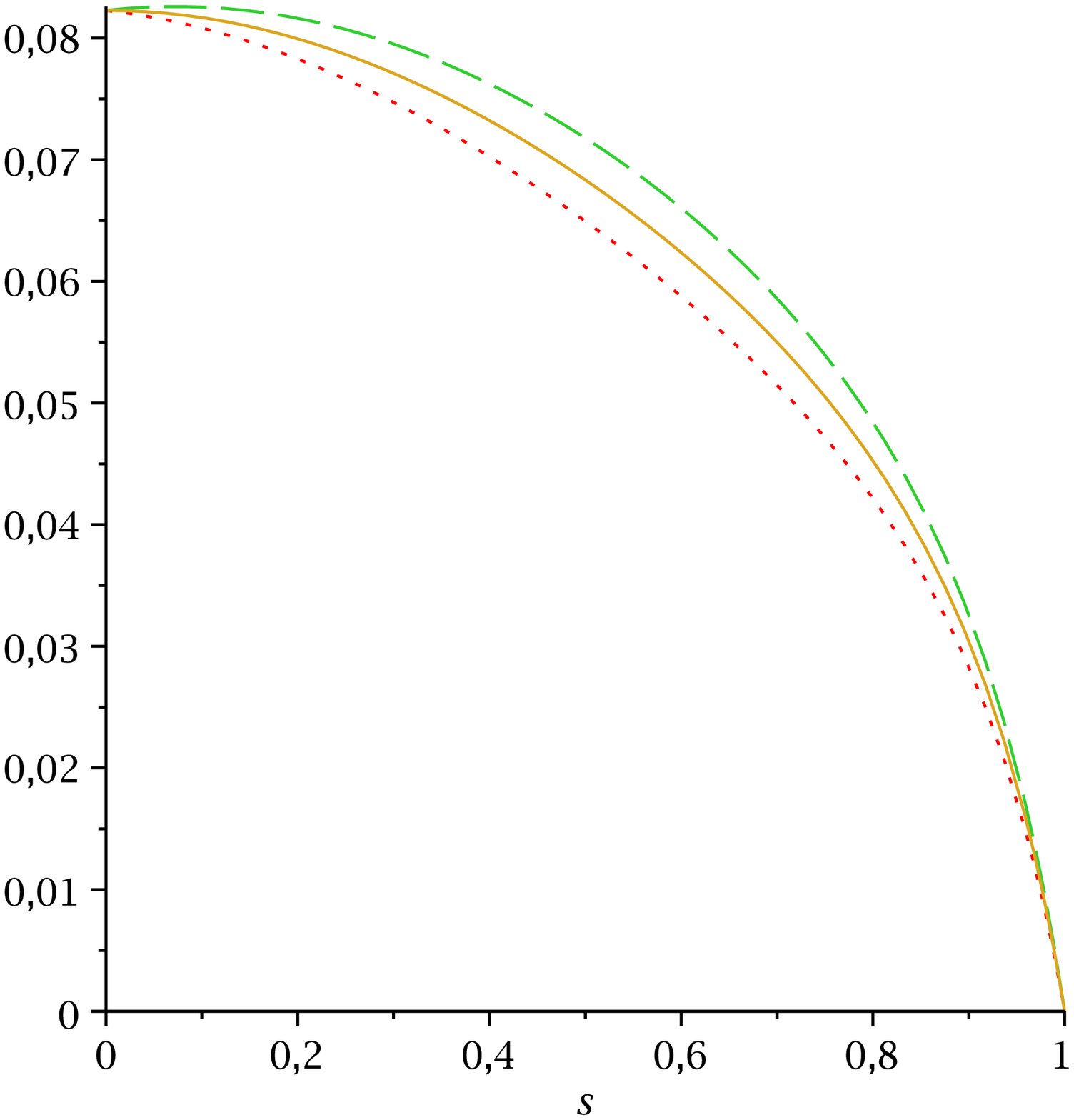}  \nonumber\\
(a) & (b) \nonumber
\end{array}
$$
\caption{\label{prr} {\it  Three profiles of  $r_{\Sigma}^2P_{rr}\equiv r_{\Sigma}^2g_{rr} T_1^1$, as function of $s$,   for  $y=0.3$ (graphic a), and  $y=1$ (graphic b) with  $\tau=2.7$, and   $q=0$, $q=0.01$ (dashes line) and $q=-0.01$ (dots line).}}
\end{figure}

Figures  4 and  5  depict the behaviour  of different energy momentum components, for a specific choice of the parameters $q$ and $\tau$. 
\begin{figure}[h]
$$
\begin{array}{cc}
 \includegraphics[scale=0.21]{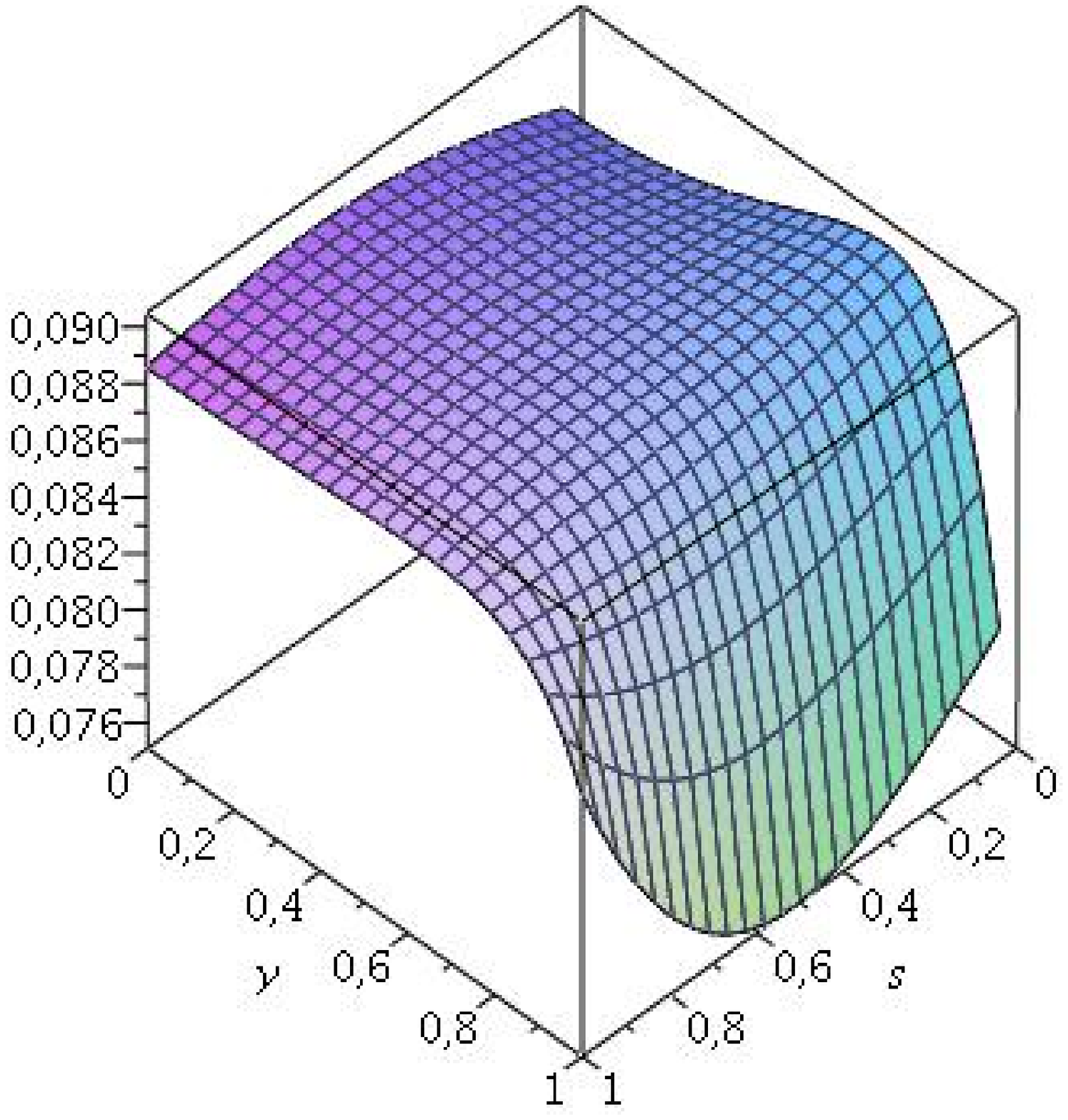}& \includegraphics[scale=0.21]{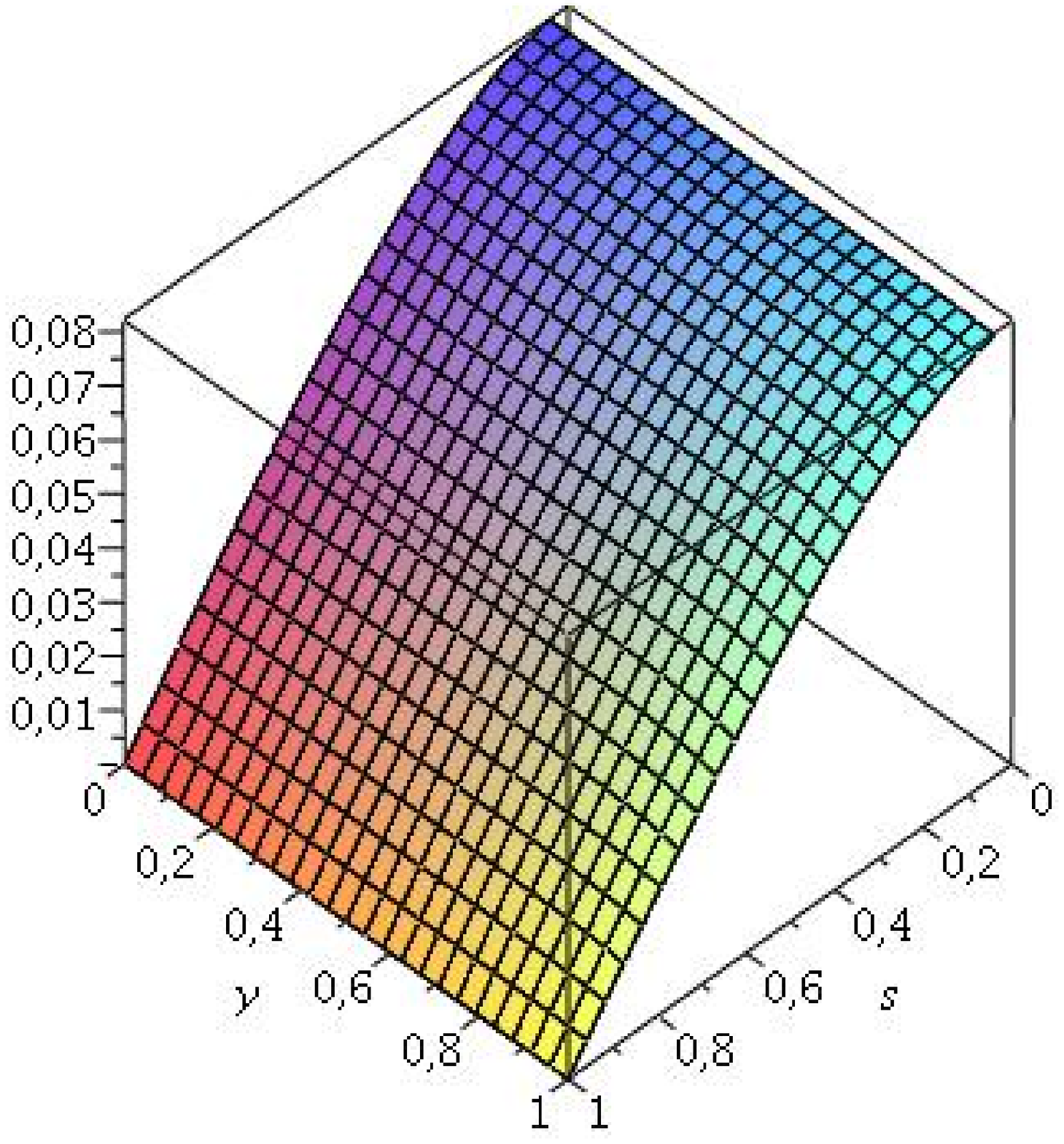}  \nonumber\\
(a) & (b) \nonumber
\end{array}
$$
\caption{\label{Ts} {\it   $-r_{\Sigma}^2T_0^0$ (graphic a), and $r_{\Sigma}^2T_1^1$  (graphic b), as functions  of  $y=\cos\theta$  and  $s$, with  $q=-0.01$ and  $\tau=2.7$.}}
\end{figure}

\begin{figure}[h]
$$
\begin{array}{cc}
 \includegraphics[scale=0.21]{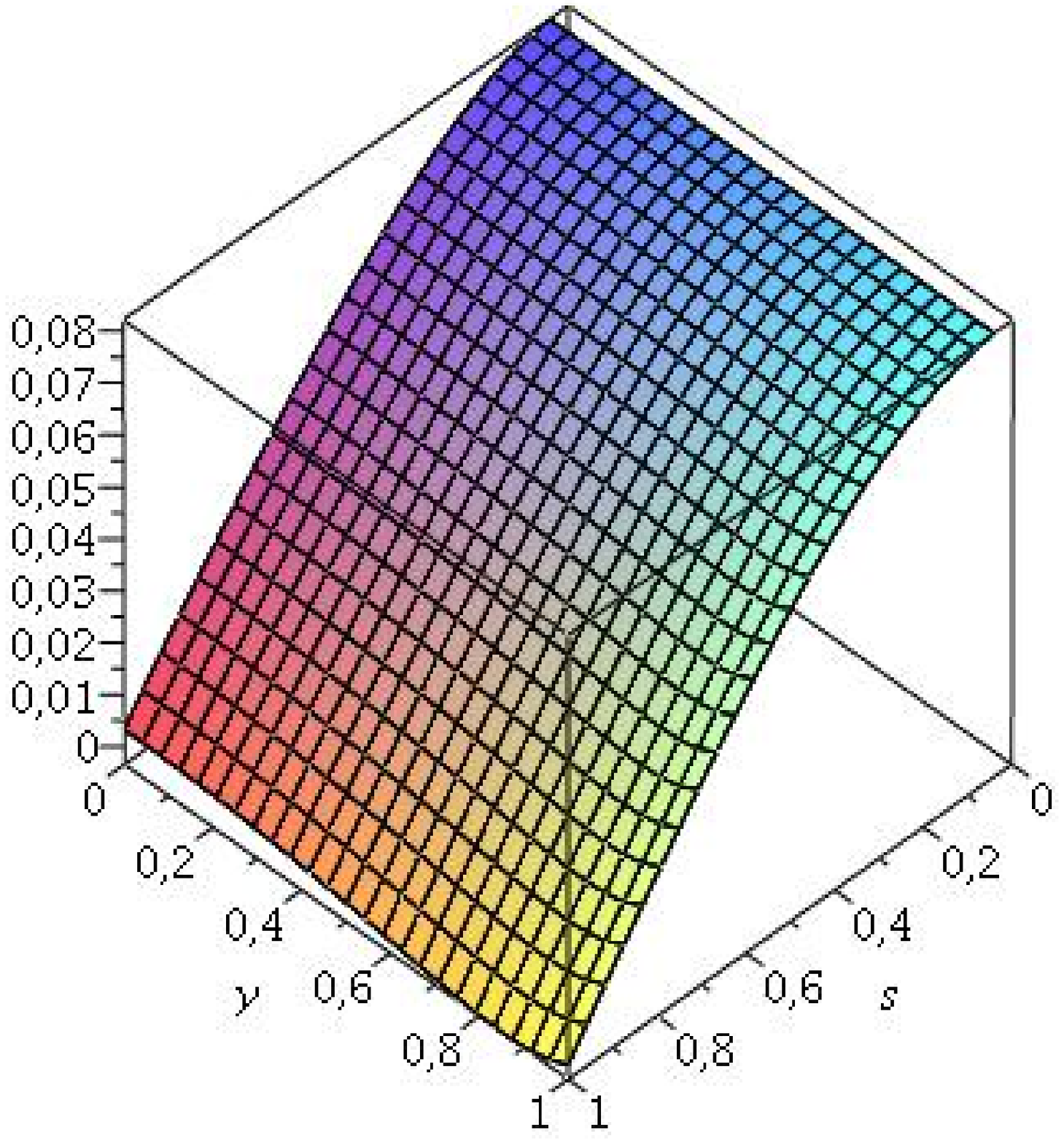}& \includegraphics[scale=0.21]{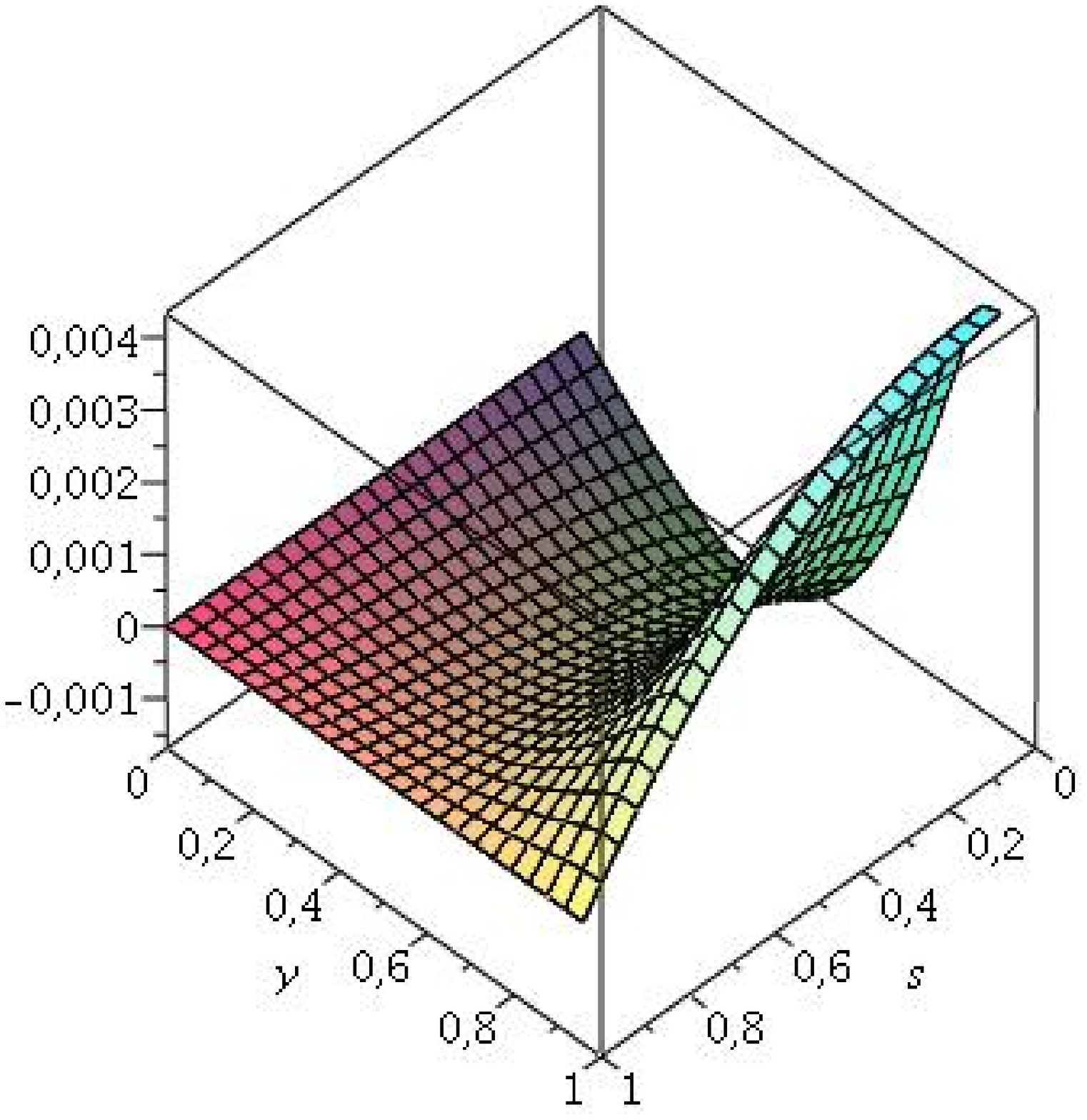}  \nonumber\\
(a) & (b) \nonumber
\end{array}
$$
\caption{\label{Ts12} {\it   (a)  $r_{\Sigma}^2T_3^3$,and (b) $r_{\Sigma}^3T_1^2$ as functions of $y$ and $s$.}}
\end{figure}

Figure 6, shows the verification of the strong energy condition $(- T_0^0)-T_1^1>0$ (also  for a specific choice of the parameters $q$ and $\tau$). 
\begin{figure}[h]
 \includegraphics[scale=0.31]{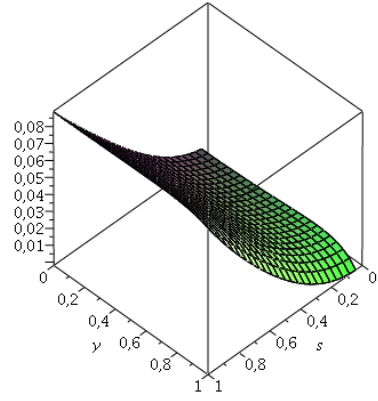}
\caption{\label{t0mt1} {\it  $r_{\Sigma}^2[(- T_0^0)-T_1^1]$ as function of   $y$ and  $s$, with $q=-0.01$ and $\tau=2.7$.}}
\end{figure}

It is instructive to depict how the mass (energy) of the source is distributed  within the volume of the compact object for a non-spherical interior (in contrast with   the perfect fluid constant density model). In figure 6 we show a density map  for positive and negative values of the parameter $q$. Brighter tone areas indicate higher energy density  values, while darker tone areas depict the opposite.  Since $M-Q^{(1)}$ solution posses axial symmetry these figures are  cut at $\phi=cte$  in cylindrical  coordinates, such that  the vertical axis ( dimensionless coordinate $z/r_{\Sigma}$) represents the symmetry axis. The figure 7 shows that the density distribution  within the source  increases along the axis and nearby the origin for $q>0$ (prolate source) whereas  it decreases along certain radial direction. However, for  the oblate source ($q<0$) the distribution is almost the opposite: the increase of density being located on the same radial direction for which  the oblate case exhibits the decrease of energy density,  whereas a decrease appears along the axis of symmetry.

\begin{figure}[ht]
$$
\begin{array}{cc}
\includegraphics[scale=0.21]{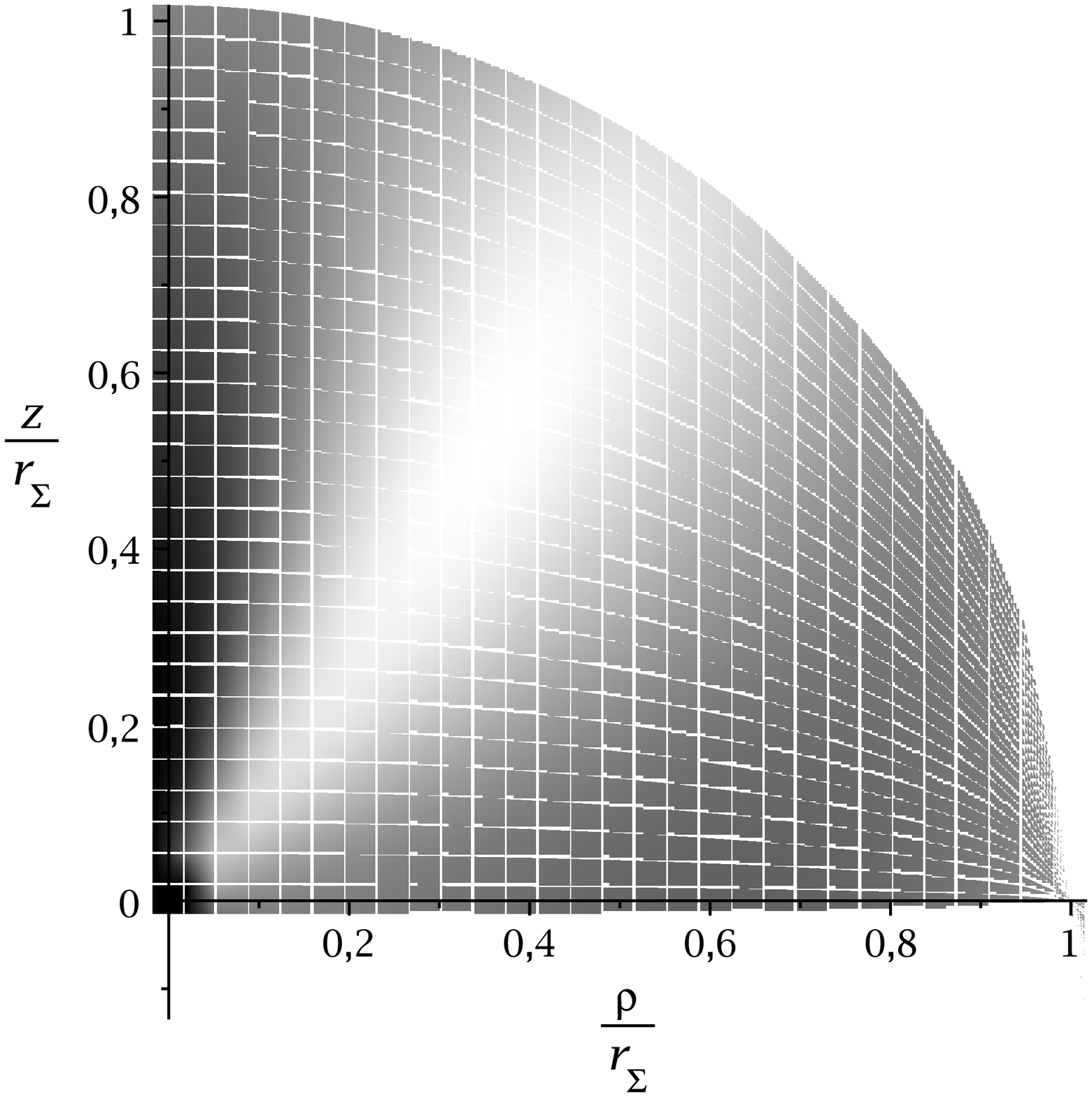}& \includegraphics[scale=0.21]{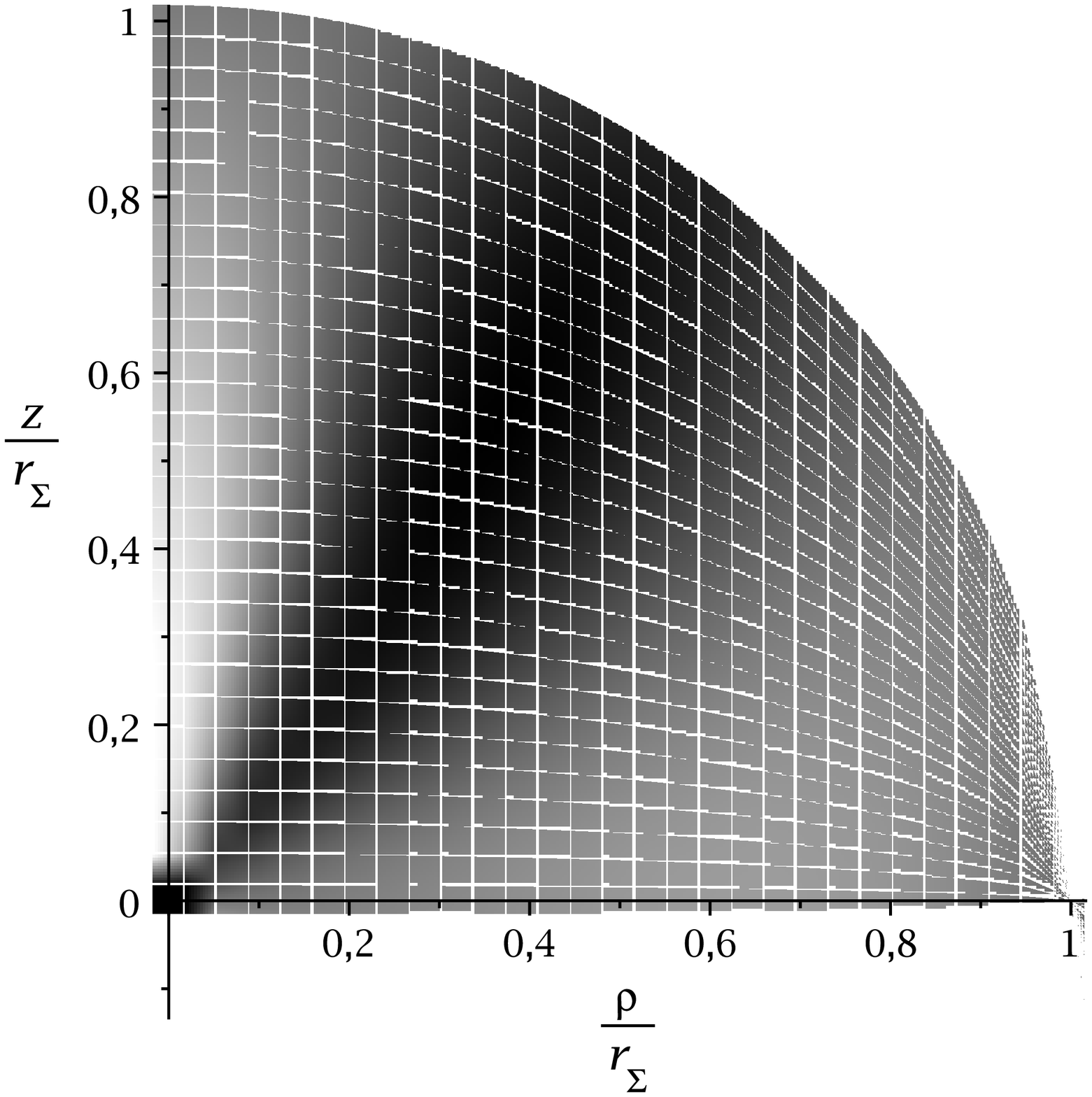}  \nonumber\\
(a) & (b) \nonumber
\end{array}
$$
\caption{\label{mapaden} {\it  Energy density distribution for    (a) $q<0$ (oblate source), and (b) $q>0$  (prolate source).}}
\end{figure}

\subsection{A source for the exterior Zipoy-Vorhees solution ($\gamma$-metric)}

The configurations to be considered here are sources of the so-called
gamma
metric ($\gamma $-metric) \cite{11}--\cite{9}. This metric, which is
also
known as Zipoy-Vorhees metric, belongs to the family of Weyl's
solutions, and is continuously linked to the Schwarzschild space--time
through one of its parameters. The interest of this metric is based on the fact that it corresponds to a solution of the
Laplace equation (in cylindrical coordinates) with the same singularity
structure (``Newtonian'' image ) as the Schwarzschild solution (a line segment \cite{9}). In
this
sense the $\gamma $-metric appears as the ``natural'' generalization of
Schwarzschild space--time to the axisymmetric case. 

In our coordinates, the exterior space--time is given by:

\begin{eqnarray}
\psi &=&\frac \gamma2 \ln\left(1-\frac{2M}{r}\right) \nonumber\\
 \Gamma &=&-\frac{\gamma^2}{2}\ln\left[\frac{(r-M)^2-M^2\cos^2\theta}{r(r-2M)}\right].
\label{gn1}
\end{eqnarray}

It is worth noticing that $\psi ,$ as given by (\ref{gn1}), corresponds
to
the Newtonian potential of a line segment of mass density $\gamma/2 $ and
length $2M,$ symmetrically distributed along the $z$ axis. The
particular
case $\gamma =1,$ corresponds to the Schwarzschild metric.

The total mass of the source is  $M_T=\gamma M,$ (because $q_0\neq1$) and the
quadrupole
moment ($M_2=qM^3$) is given by
\begin{equation}
M_2=\frac \gamma 3M^3\left( 1-\gamma ^2\right) .  \label{9}
\end{equation}
So that $\gamma >1$ ($\gamma <1$) corresponds to an oblate (prolate)
spheroid.

Then, following the algorithm described above, we may write for the interior of the $\gamma$ metric:
\begin{eqnarray}
\hat \psi_{\Sigma}\equiv \psi_{\Sigma}-\psi^s_{\Sigma}=\frac h2 \ln\left(1-\frac{2}{\tau}\right)\nonumber\\
\hat \Gamma_{\Sigma}\equiv \Gamma_{\Sigma}-\Gamma^s_{\Sigma}=-\left(h+\frac{h^2}{2}\right) \ln\left[\frac{(\tau-1)^2-\cos^2\theta}{\tau(\tau-2)}\right]
\label{fmg}
\end{eqnarray}
with $\gamma=1+h$, $h \neq 0$.
From the expressions above we obtain,  using (\ref{funcint})
\begin{equation}
\hat a(s,y)=-\frac{h s^2}{2}\left[ -2\frac{s-1}{\tau -2}+(2s-3)\ln\left(\frac{\tau -2}{\tau}\right)\right].
\label{hata}
\end{equation}
\begin{widetext}
\begin{equation}
\hat g(s,y)=-s^3(h+\frac{h^2}{2})\left\{ \frac{2(s-1)(1-\tau)(1-y^2)}{(\tau -2)[(\tau-1)^2-y^2]}+(4-3s)\ln\left[\frac{(\tau -1)^2-y^2}{(\tau)(\tau-2)}\right]\right\}.
\label{hatg}
\end{equation}
\end{widetext}

We can now check that the total mass of any of the configurations serving as sources of the $\gamma$ metric equals the monopole ($M_0$)  of the exterior metric, i.e. $q_0=\gamma M=M_0$.

Indeed, from (\ref{m1}) and (\ref{fmg})
\begin{equation}
\hat{\psi}^{\prime}_{\Sigma}=
\psi^{\prime}_{\Sigma}-(\psi^s)^{\prime}_{\Sigma}=\frac{(\gamma-1)}{r_{\Sigma} (\tau-2)}-\sum_{n=1}^{\infty}q_nQ^{\prime}_n(r_{\Sigma})P_n(y).
\label{m11}
\end{equation}
Then, following the same orthogonality arguments about Legendre polynomials used before we obtain

\begin{equation}
M_T=M+\frac{h}{2}r_{\Sigma}^2 A_{\Sigma}\int_0^{\pi}d\theta \sin\theta \frac{M}{r_\Sigma(r_\Sigma-2M)}=\gamma M.
\label{masatotalB}
\end{equation}

Also, as in the previous example, we can calculate the proper length along the symmetry axis, as well as the proper equatorial radius.

From these calculations, a strange, and, apparently contradictory result, appears. Indeed, our calculations show (see for example figure 9), that  no matter the value of $\tau$, the functions $e^{-\hat a(y=1)}$ and $e^{(\hat g-\hat a)(y=0)}$  are such, that $h<0, (\gamma<1)$ implies $l_z<l_{\rho}$ (oblate) whereas $h>0, (\gamma>1)$ produces a prolate source  since  $l_z>l_{\rho}$. This of course is at variance with (\ref{9}), from which exactly the opposite is obtained, i.e.  $h<0, (\gamma<1)$ implies that the source is prolate, whereas $h>0, (\gamma>1)$ produces a oblate source. 
 This result reflects itself in the ellipticity of the source, which  for this case, is positive (prolate source)  when $h>0$, instead of the criterion (\ref{9}), as can be seen in figure 8.

\begin{figure}[h]
 \includegraphics[scale=0.31]{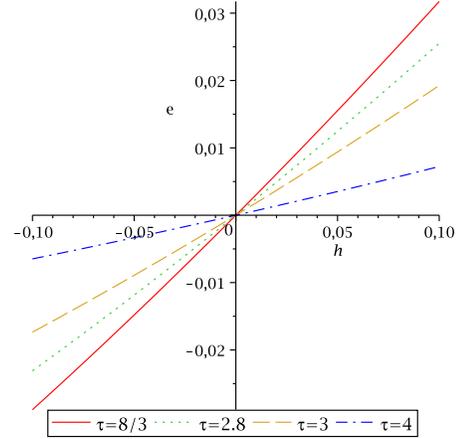}  
\caption{\label{ellip} {\it  Relation between the ellipticity of the source $e$ and its  parameter $h$ for different values of $\tau$.}}
\end{figure}

This strange ``anomaly''  also shows up  from the evaluation of the total area of the boundary surface. Indeed, a simple calculation using  (\ref{area}), shows that for the source of the $\gamma-$metric, independently of its compactness (any value of the parameter $\tau$),  we obtain that for any positive value of $h$ ($\gamma>1$) (oblate source), the area of the boundary surface behaves as expected,  since $\hat \Gamma_{\Sigma}-2\hat\psi_{\Sigma} >0$, however, for  any negative value of the parameter $h$ ($\gamma<1$) (prolate source), the  value of the surface area is smaller than the corresponding  to  the spherical case. 

At this point it should be pointed out, that a similar result was already reported by  Bonnor \cite{r3},  with respect to an interior solution for the Curzon space--time.

But, what is the origin of such a strange result?  Before trying to answer  to such a question, let us calculate  the length of the meridian curves $l_{\theta}$ ($r=r_{\Sigma}$, $\phi=$cte) and the circular equator $l_{\phi}$ ($\theta=\pi/2$, $r=r_{\Sigma}$). Obviously $l_{\theta}> l_{\phi}$, would indicate a prolate source, whereas an oblate source is expected if  $l_{\theta}< l_{\phi}$:
Thus, we have 
\begin{equation}
l_{\theta}=2 r_{\Sigma} \int_0^{\pi} e^{(\hat g-\hat a)_{\Sigma}} \  d \theta \ , \  l_{\phi}= r_{\Sigma} \int_0^{2\pi} e^{-\hat a_{\Sigma}(\theta=\pi/2)} \  d \phi.
\label{lmeriequa}
\end{equation}
For the spherical case ($\hat a=\hat g=0$) both lengths are equal $l_{\theta}=l_{\phi}=2\pi r_{\Sigma}$.

Instead, in the general non-spherical case we have that 
\begin{equation}
l_{\theta}=2 r_{\Sigma} \int_0^{\pi} e^{(\hat{\Gamma}-\hat {\Psi})_{\Sigma}} \  d \theta \ , \  l_{\phi}=2 \pi r_{\Sigma}  e^{-\hat{\Psi} _{\Sigma}(\theta=\pi/2)} 
\label{lmeriequa2}
\end{equation}

Now, if we compare the lengths of meridian curves and circular equator (\ref{lmeriequa}) for this case, we have  from (\ref{hata}-\ref{hatg}) that
\begin{equation}
l_{\theta}=l_{\phi} \frac{1}{\pi} \int_0^{\pi}\left(\frac{c_1}{c_1+\sin^2\theta}\right)^{c_2} \ d \theta
\end{equation}
where  the proper  equator curve length is  $\displaystyle{l_{\phi}=2\pi r_{\Sigma}\left(\frac{\tau}{\tau-2}\right)^{h/2}}$ and $c_1=\tau(\tau-2)$, $\displaystyle{c_2=h+\frac{h^2}{2}=\frac 12(\gamma^2-1)}$. 

Therefore, if $\gamma>1$ ($\gamma<1$) then $c_2>0$ ($c_2<0$) and we get $l_{\theta}<l_{\phi}$ oblate source  ($l_{\theta}>l_{\phi}$ prolate source) since $\displaystyle{\frac{c_1}{c_1+\sin^2\theta}<1}$. All this in full agreement with the conclusions extracted from (\ref{9}).

At this point, it is instructive to get back to the example presented in section III.B (the source for the  $M-Q^{(1)}$ solution). In that case,  the obtained expressions  for the proper length along the symmetry axis and  the proper equatorial radius, lead to the expected criterion to elucidate whether the source is oblate or prolate. So, it is legitimate to ask: What do we obtain, if we calculate for $M-Q^{(1)}$ source, the meridian and equator curves lengths on the surface? As we shall see, in this case the result is the same as the one obtained from the calculation of the the proper length along the symmetry axis and  the proper equatorial radius, i.e. the  contradictory result appearing for the source of the $\gamma$ metric  and that of the Curzon metric, does not appear for the source of the $M-Q^{(1)}$ solution.

 Indeed, for the  $M-Q^{(1)}$ solution we have from (\ref{psimq1})  that 
\begin{equation}
l_{\theta}=l_{\phi} \frac{1}{\pi} \int_0^{\pi}e^{\hat{\Gamma}_{\Sigma}} e^{\hat{\Psi}_{\Sigma}(y=0)-\hat{\Psi}_{\Sigma}} \ d \theta
\label{relalong}
\end{equation}
where $\hat{\Gamma}_{\Sigma}$ is given in (\ref{aygsimple}) and 
\begin{widetext}
\begin{equation}
 e^{\hat{\Psi}_{\Sigma}(y=0)-\hat{\Psi}_{\Sigma}}=\left(1-\frac{2}{\tau}\right)^{-\frac{15}{32}y^2 q(3\tau^2-6\tau+7)}\exp\left[q\left( -\frac{45}{16}y^2(\tau-1)+\frac 54 \frac{\tau -1}{(\tau-1)^2-y^2}-\frac{5}{4(\tau -1)}\right)\right]
\label{esto}
\end{equation}
\end{widetext}

Due to the cumbersome of the expression (\ref{relalong}) for this case, we shall resort to a numerical evaluation of the integral. The result is  that independently on  what the absolute values of $\tau$ and $q$ are, we have that $q<0$ implies $l_{\theta}<l_{\phi}$ and hence the source is oblate, and inversely if $q>0$, then $l_{\theta}>l_{\phi}$, and hence the source is prolate,  as expected.

After all these calculations and comments, we have not a definitive explanation about the  origin of the anomaly mentioned above with respect to the $\gamma$ (and the Curzon) metric.  
\begin{figure}[h]
$$
\begin{array}{cc}
 \includegraphics[scale=0.21]{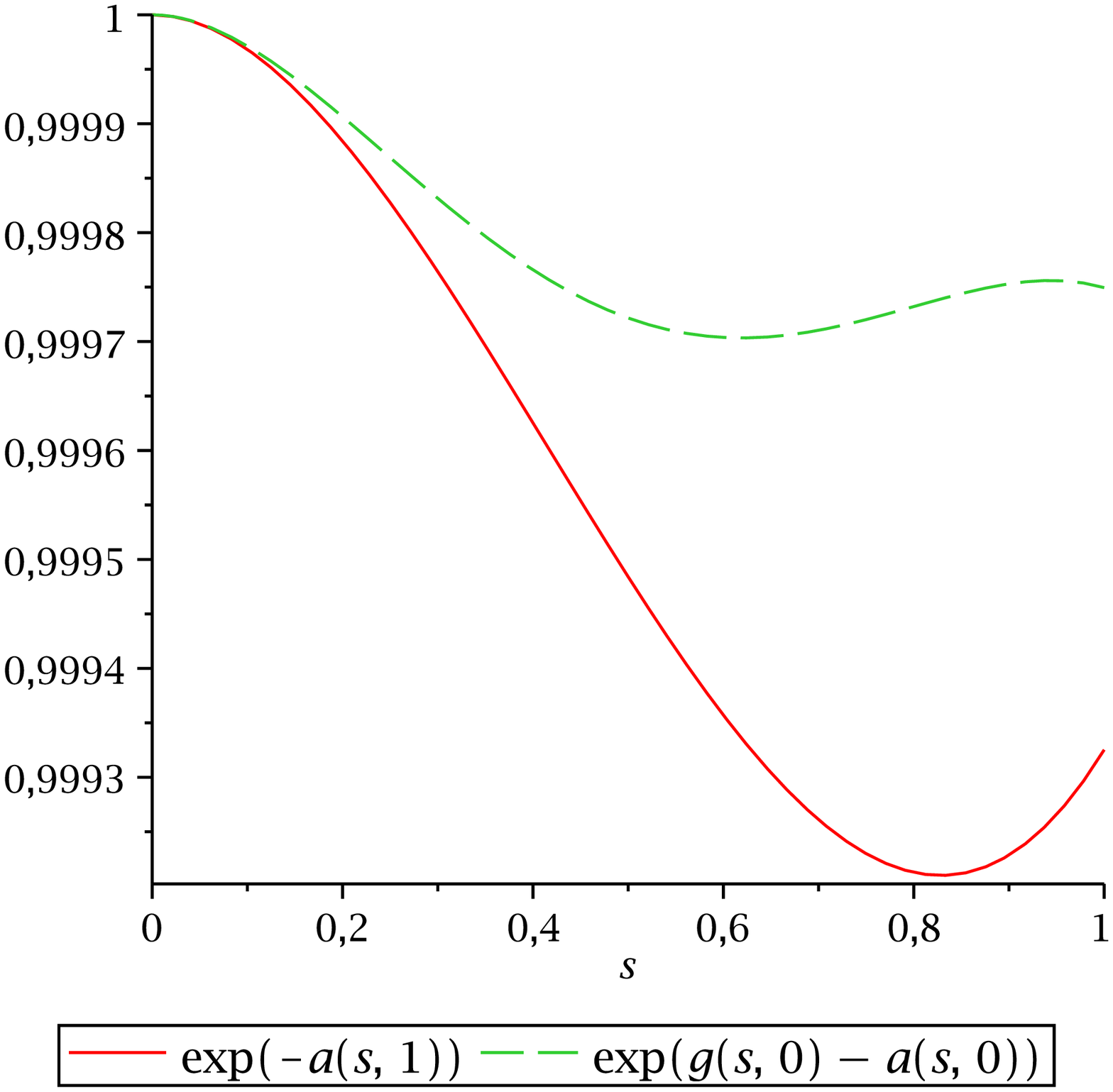}& \includegraphics[scale=0.21]{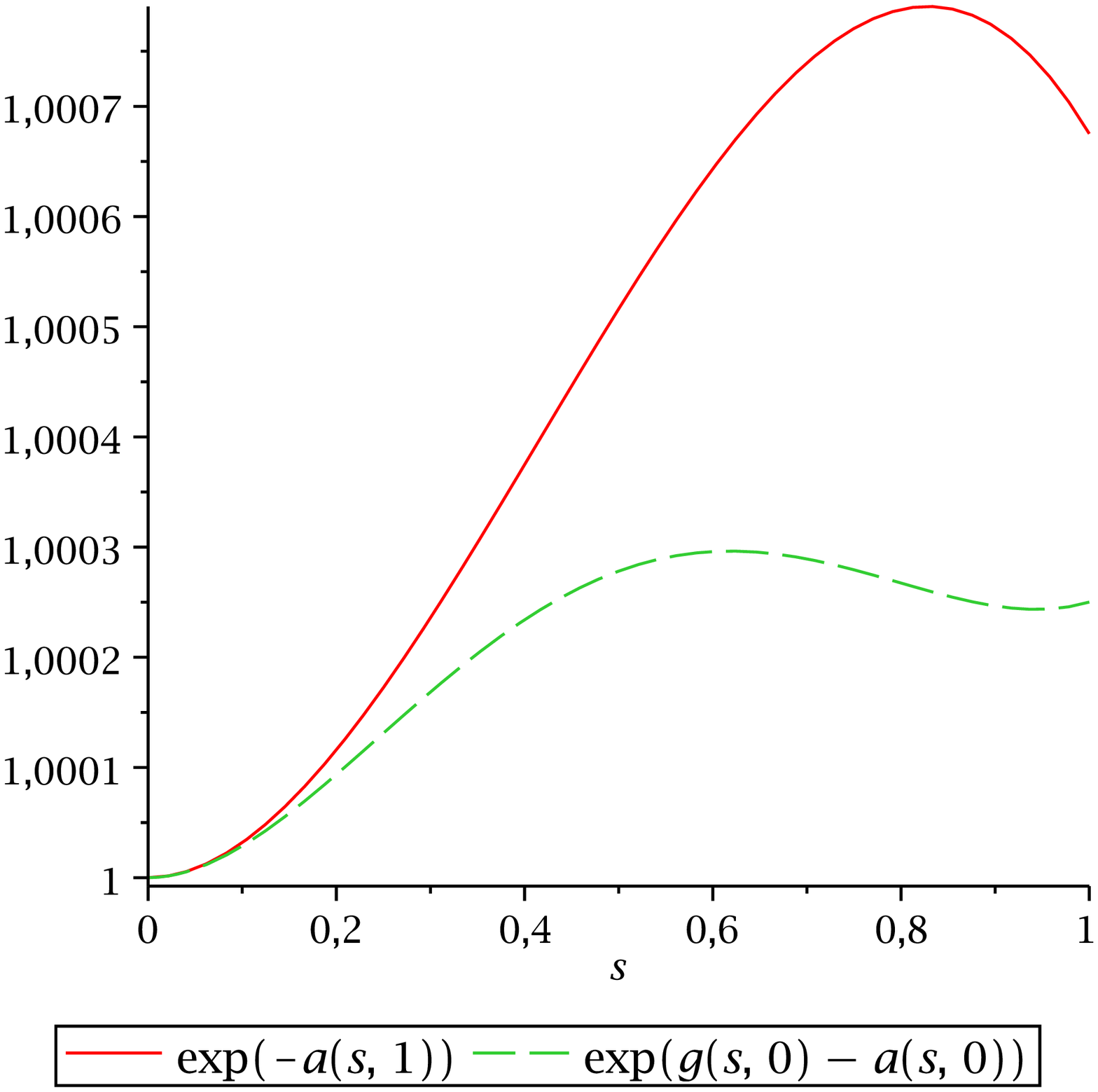}  \nonumber\\
(a) & (b) \nonumber
\end{array}
$$
\caption{\label{longitudes} {\it  Functions  $exp(\hat g(s,y=0)-\hat a(s,y=0))$ and  $exp(-\hat a(s,y=1))$  for positive (curve b) and negative (curve a) values of the  parameter $h=\pm 0.001$ and   $\tau=2.7$.}}
\end{figure}

The regularity of the metric variables as well of the energy--momentum components, is assured by the choice of functions $\hat a$ and $\hat g$, whereas 
the fulfillment of the energy conditions depends on the assumed values of the parameters $h$ and $\tau$. Thus, assuming for $\tau$ a value close to the  minimum allowed in the spherically symmetric case, i.e.  $\tau>8/3$,  it can be shown that $-T_0^0>0$ (positive energy density) and $(-T_0^0)-T_1^1>0$ (strong energy condition), for different values of $h$ are satisfied, depending on whether the source is oblate or prolate. Indeed, for a prolate source ($h<0$), the well behaviour of the physical variables is assured for values of $h$ of the order of $-0.1$. Instead, for an oblate source, physically meaningful sources require values of $h$ no greater than $h=0.001$, for otherwise, different kind of pathologies appear.
Finally, we notice that the density profile (with respect to $r$) depends on the sign of $h$. Thus, for the prolate case ($h<0$) $\frac{\partial}{\partial r}(-T_0^0) <0$ within the source, whereas the opposite happens in the oblate case, as indicated in figure 10. Since positive gradients of energy density would imply instability, these latter models could only be used as initial configurations in a collapse scenario.

\begin{figure}[h]
$$
\begin{array}{cc}
 \includegraphics[scale=0.21]{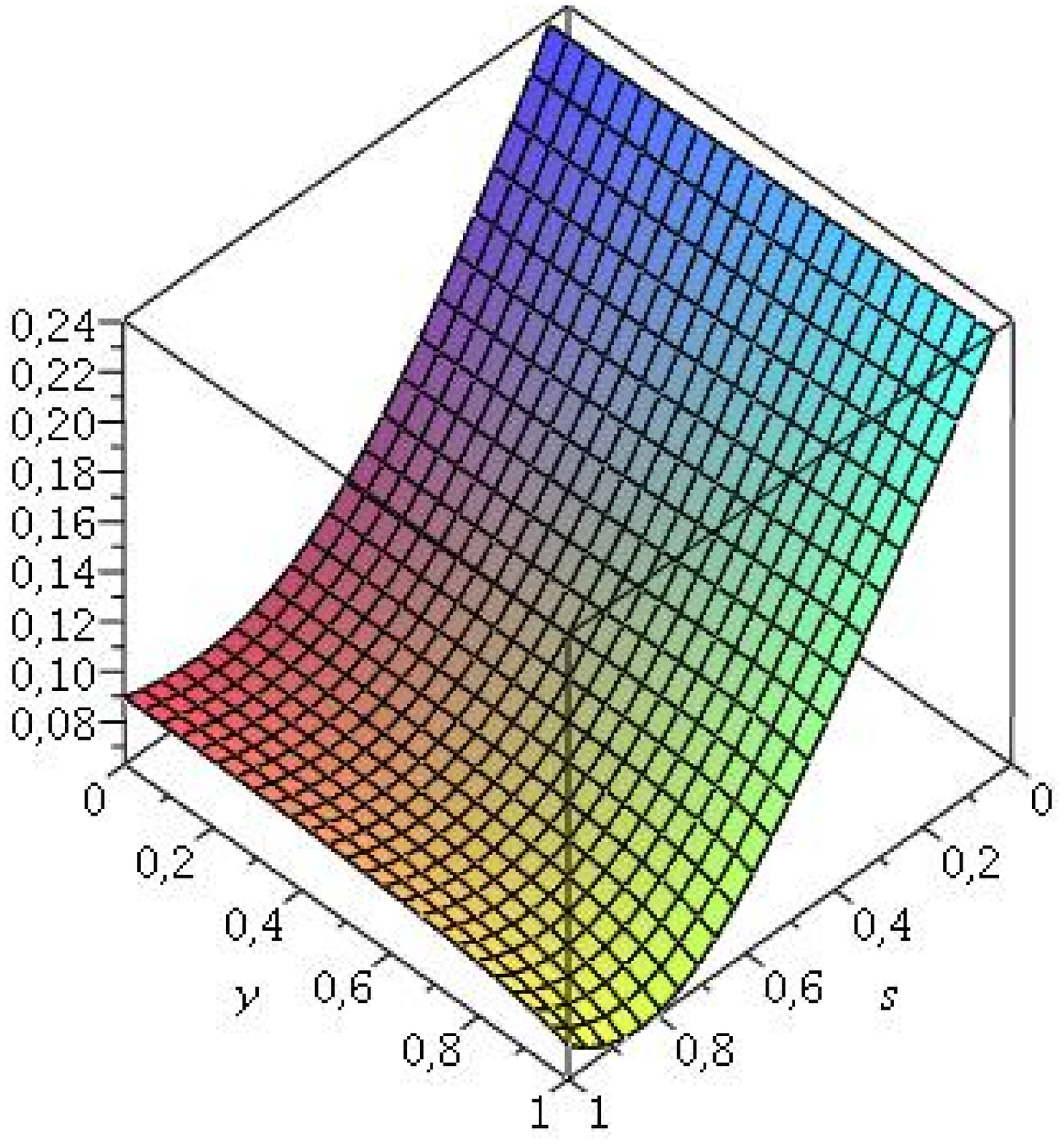}& \includegraphics[scale=0.21]{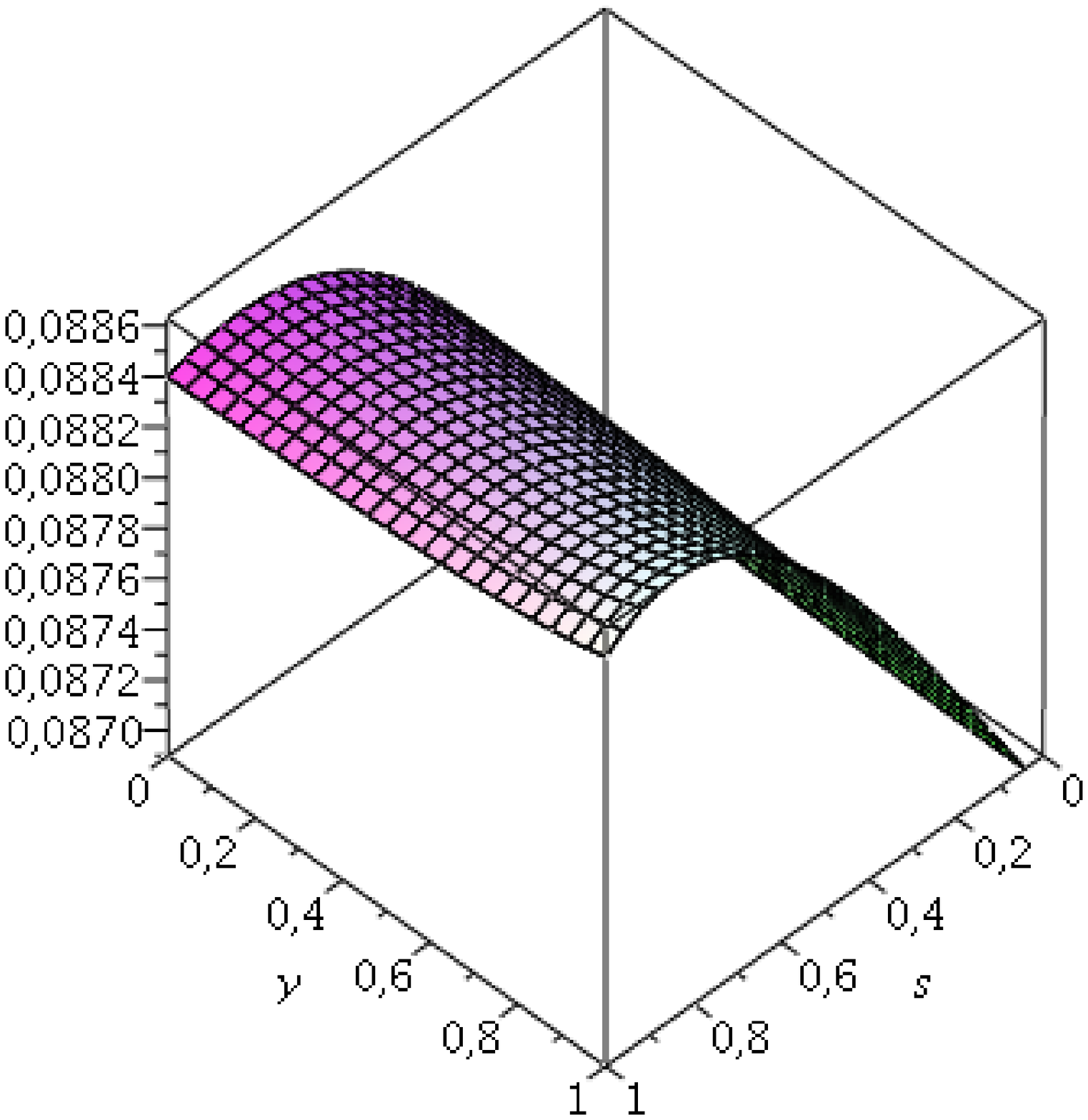}  \nonumber\\
(a) & (b) \nonumber
\end{array}
$$
\caption{\label{T00zv} {\it   $-r_{\Sigma}^2T_0^0$ as function of $s$ and $y$, for $h=0.001$ (figure b), and $h=-0.1$, (figure a), and  $\tau=2.7$. }}
\end{figure}

Figures  11 and  12  depict the behaviour  of different energy momentum components, for a specific choice of the parameters $h$ and $\tau$, whereas figure 13, shows the verification of the strong energy condition $(- T_0^0)-T_1^1>0$ for the same choice of the parameters: 
\begin{figure}[h]
$$
\begin{array}{cc}
 \includegraphics[scale=0.21]{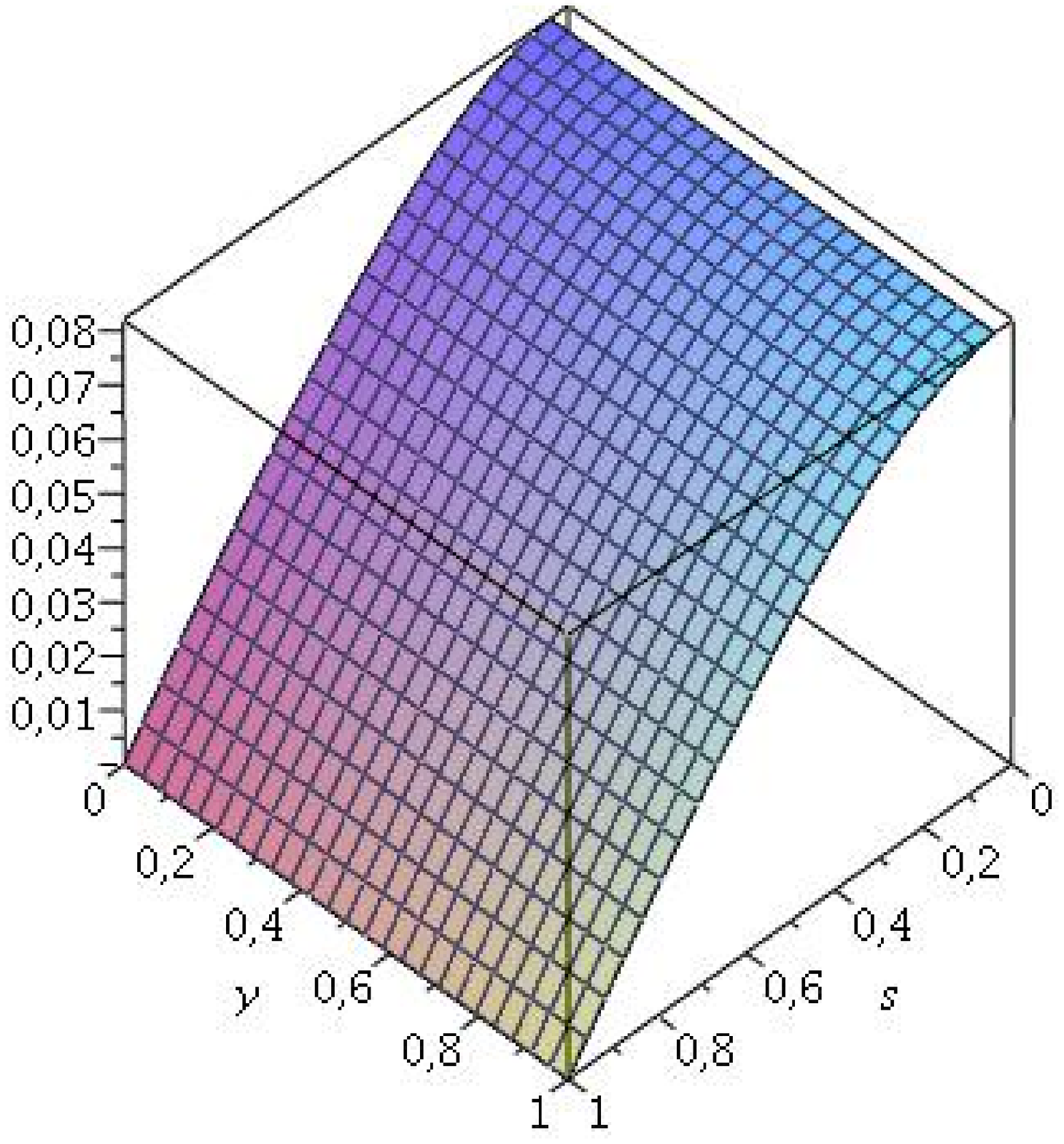}& \includegraphics[scale=0.21]{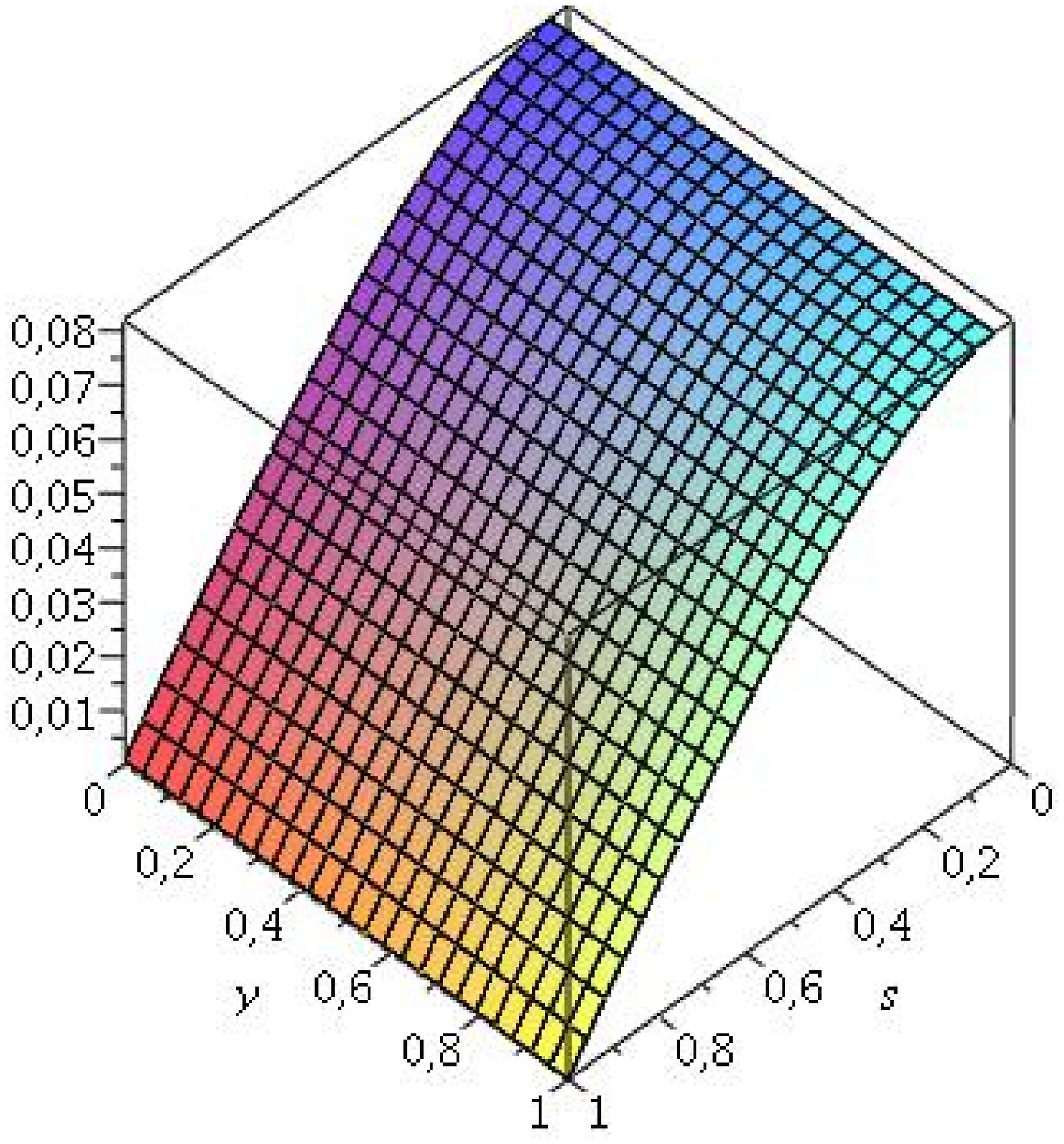}  \nonumber\\
(a) & (b) \nonumber
\end{array}
$$
\caption{\label{Tszv} {\it   $-r_{\Sigma}^2T_1^1$ (graphic a), and $r_{\Sigma}^2T_3^3$  (graphic b), as functions  of  $y=\cos\theta$  and  $s$, with  $h=0.001$ and  $\tau=2.7$.}}
\end{figure}

\begin{figure}[h]
$$
 \includegraphics[scale=0.21]{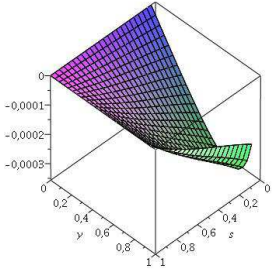} 
$$
\caption{\label{Ts12ZV} {\it  $r_{\Sigma}^3T_1^2$ as function of $y$ and $s$.}}
\end{figure}

\begin{figure}[h]
 \includegraphics[scale=0.31]{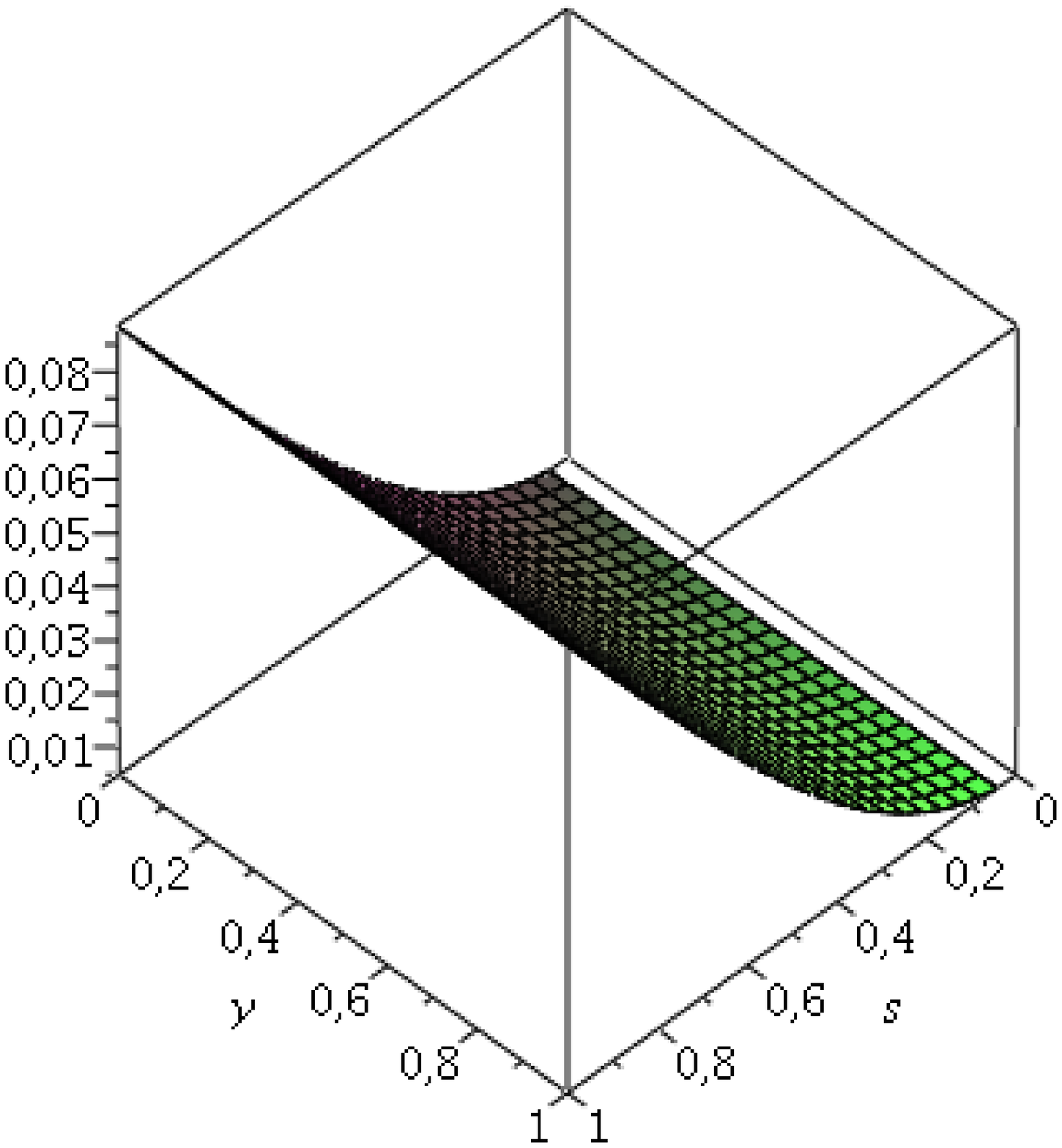}
\caption{\label{t0mt1ZV} {\it  $r_{\Sigma}^2[(- T_0^0)-T_1^1]$ as function of   $y$ and  $s$, with $h=0.001$ and $\tau=2.7$.}}
\end{figure}

 With respect to the intrinsically non-spherical component $T_1^2$, another difference arises between the source of the  $\gamma$ metric and the source of the $M-Q^{(1)}$ solution. Indeed, since the metric function  $\hat a$ (\ref{hata}) does not depend on the angular variable, the component $T_1^2$ from (\ref{t12}) does not contain quadratic terms  in the quadrupole moment and can be explicitly expressed in terms of that multipole moment as follows
\begin{widetext}
\begin{eqnarray}
T_1^2&=&\frac{\kappa }{r_{\Sigma}^3} \frac{3q}{2 \gamma}  \left\lbrace 3 \frac{\cos\theta}{\sin\theta} \left[\frac{2(1-\tau)(1-y^2)(4s-3)}{3(\tau-2)((\tau-1)^2-y^2)}+4(1-s)\ln\left(\frac{(\tau-1)^2-y^2}{\tau(\tau-2)}\right) \right]+\right.\nonumber\\
&+&\left. \frac{2y \sqrt{1-y^2}}{(\tau-1)^2-y^2}\left[1+\frac{2s^2}{\sqrt{\tau-2s^2}(3\sqrt{\tau-2}-\sqrt{\tau-2s^2})}\right]\left[
(4-3s)+\frac{2\tau(s-1)(1-\tau)}{(\tau-1)^2-y^2} \right]\right\rbrace
\end{eqnarray}
\end{widetext}
where the fact that  $\displaystyle{h+\frac{h^2}{2}=-\frac{3 q}{2 \gamma}}$ has been taken into account, $q\equiv M_2/M^3$, being the quadrupole parameter of the $\gamma$ metric.

On the other hand, the $M-Q^{(1)}$ metric contains terms which are proportional to the quadrupole parameter, and other which are quadratic in that parameter. For simplicity, we shall assume slight deviations from sphericity, implying that we can safely neglect terms of order   $q^2$. Hence we obtain from (\ref{prepsimq1}) and (\ref{t12}) the following expression for the source of the  $M-Q^{(1)}$ solution
\begin{widetext}
\begin{eqnarray}
T_1^2&=&\frac{\kappa }{r_{\Sigma}^3} q  \left\lbrace \frac{\cos\theta}{\sin\theta} \left[-12 \Gamma_q (s-1)+r_{\Sigma} \Gamma^{\prime}_q (4s-3)+\dot \Gamma_q(4-3s)+r_{\Sigma} (\dot \Gamma_q)^{\prime} (s-1)\right]+\right.\nonumber\\
&+&\left. \frac{2s}{\sqrt{\tau-2s^2}(3\sqrt{\tau-2}-\sqrt{\tau-2s^2})}\left[s\dot \Gamma_q (4-3s)+s r_{\Sigma}(\dot \Gamma_q)^{\prime} (s-1)+\dot \psi_q(4s-6)+2r_{\Sigma}(\dot \psi_q)^{\prime} (1-s) \right]\right\rbrace+O(q^2).\nonumber\\
\end{eqnarray}
\end{widetext}

\section{Discussion}
We have put forward a general method to find sources of Weyl metrics. The method includes the enforcement of the junction conditions on the boundary surface, which allows to obtain global solutions. On the other hand basic physical requirements are incorporated, so that all possibles sources are characterized by physically meaningful variables.
This is achieved, by adjusting appropriately,  the ansatz used for the obtention of the interior metric functions. Here we only considered the simplest case ($\mathbb{F}=\mathbb{G}=0$), of course our method allows for dealing with more general situations.

To illustrate the method, we have found sources for two  well known  space--times  of the Weyl family, namely: the M-Q$^{(1)}$ and the $\gamma$ metrics. 

As one of the most interesting aspects of the obtained models, we would like to stress the fact that we are able to link the physical variables describing the source (its structure) with an observable quantity such as the quadrupole moment. This may pave the way to infer the equation of state of the source, from observations of the gravitational field produced by  such a source.

It should be pointed out, that another source for the M-Q$^{(1)}$ was found in \cite{7}, however in that example, it was necessary to introduce a spherical core, in order to
assure acceptable physical behaviour at the centre. It consists of a sphere of incompressible
fluid with positive energy density, larger than pressure and which
matches smoothly to the outer part of the source. Such a, rather artificial, trick, is unnecessary in the example presented here, which for a wide range of the parameters presents an acceptable physical behaviour.

Also, a source for the $\gamma$ metric was obtained in \cite{6}, however the resulting interior metric in that reference,   is restricted by the Weyl gauge, which as already mentioned in the Introduction, may represent a too stringent condition.

A natural question arises at this point, namely: what is the  range of values of $\tau$ and $q$ (or $h$ in the case of the source of the $\gamma$ metric), for which our models exhibit acceptable physical properties?. To answer to such a question we have run a large number of models to establish  that range. We have focused  on the fulfillment of  Positive Energy Density (P.E.D.) ($-T_0^0>0$), Strong Energy Condition (S.E.C.) ($(-T_0^0)-T_i^i>0$) and Positive Radial Pressure (P.R.P.) ($P_{rr}=g_{11}T_1^1>0$).The results of this  search are  shown in Tables I--III.  

As  expected, and as it is apparent from tables I and II,  in the case of the source of the $M-Q^{(1)}$ solution,  the greater is  the compactness (the smaller is $\tau$) of the source, the larger are the values (the absolute values) of $q$ required to satisfy the above mentioned physical conditions. Thus,  close to  the minimum value of $\tau$ (maximal compactness) ($\tau=2.7$), we need values of  $q\geq -0.005$ (oblate source) and $q\leq 0.01$ (prolate source) to guarantee  a good physical behaviour.

The situation is similar for the source of the $\gamma$ metric (table III). In this case  we need $h\geq-0.01$ (prolate source) and $h<0.001$ (oblate source) to produce physically meaningful models.

There exists a large set of physical scenarios, from neutron stars to white dwarfs \cite{mexico}, characterized by parameters within  these ranges of values. For  values of $\tau$ close  to  $8/3$, we must resort to  neutron stars. Indeed,  a typical neutron star of $M=1.8 M_{\odot}$ and radius $R\sim 10$ km leads to $\tau\sim 3.76$ and a more massive star of $M=5 M_{\odot}$ with  radius $R=20$ km $\sim 2.87 \cdot 10^{-5} R_{\odot}$ leads to $\tau\sim 2.71$. On the other hand the mass of a white dwarf is bounded by the Chandrasekar limit \cite{chandra} $M=1.4 M_{\odot}$, so a typical white dwarf of that mass and a radius of order the terrestrial radius $R\sim 10^{-2} R_{\odot}$ gives $\tau\sim 3.08 \cdot 10^{3}$. 

In \cite{laarakers} several numerical models of neutron stars are constructed for different  equations of state (EOS) with values of the  gravitational mass in the interval between $1.0$ and $1.8$ solar masses, and  different  values of the quadrupole moment $q$. Thus, for $M=1.8 M_{\odot}$ the models cover a range of values of $q$ in the interval $(-0.024, -2.6)$ depending on the EOS used.  For  one solar mass  object, the  quadrupole moment is in the range $(-0.158, -5.63)$. These estimates are in perfect agreement with the values of our tables. Indeed,  models with  a quadrupole parameter $q=-0.01$ and  $\tau\geq 2.8$, correspond to a source with good physical behaviour.

In \cite{MNRAS} a  comprehensive  study of the quadrupole moment of neutron stars and strange stars was carried out. These authors show that for static non--rotating configurations, the existence of extremely compact objects with $\tau \sim 3$ is allowed by some equations of state \cite{gandolfi2010}.

 A typical neutron star  with mass $M=1.4 M_{\odot}$ corresponds to a range of values of  $\tau$, between $\sim 4.8$ to $\sim 7.0$. Assuming the  value of the  mass of neutron stars to be around $1.25 M_{\odot}$, then the corresponding value of $\tau$  is about $8$.

 Thus,  the good physical behaviour of  our models  with these values of $\tau$ is assured for  a value of $q$ not necessarily small (see tables I-II). In fact, these authors in \cite{MNRAS} estimate the interval for possible values of $q$ for the  astrophysically relevant neutron--star models. These estimates are related to the Kerr factor $\tilde{q}\equiv M_2 M/J^2$, $J$ being the angular momentum, and the dimensionless parameter $j\equiv J/M^2$ is considered to be in the interval $0.2 - 0.5$, and our quadrupole parameter can be expressed as $q=\tilde{q} j^2$. Thus, the resulting values range  from  $q\sim 0.06$ for the most extreme objects close to maximal mass, up to $q\sim 0.36$ for low mass objects (a value for $j=0.2$ is considered).

A more recent accurate measurement of a large neutron
star mass in the system J1614-2230, provides a mass $M=1.97\pm 0.04 M_{\odot}$ \cite{gandolfi}. 
Attempts to infer neutron star radii  have favored relatively small values ranging from $9$ to $12$ km \cite{gandolfitherein}- \cite{gandolfitherein2}. These values implies a  $\tau$ equal to  $3.09$ and $4.12$ respectively.

Finally, it is worth noticing that positive pressure gradients, which imply instability in the perfect fluid case, does not necessarily imply instability in our case, due to the anisotropy of pressure, which is the source of other ``force'' terms, besides the pressure gradients (see equations (21) and (22) in \cite{H1}).
\begin{widetext}

\begin{table}[htdp]
\caption{Fulfillment (F) or violation (V) of different  criteria for good physical behaviour: Positive Energy Density (P.E.D.) ($-T_0^0>0$), Strong Energy Condition (S.E.C.) ($(-T_0^0)-T_i^i>0$) and Positive Radial Pressure (P.R.P.) ($P_{rr}=g_{11}T_1^1>0$). The symbol $^*$ over F means  that although the criterion is fulfilled, nevertheless $\partial_r P_{rr}$ changes its sign in the interval $s \in [0,1]$ within the source.
This table corresponds to the $M-Q^{(1)}$ solution for negative values of $q$ and a sequence of different values of the parameter $\tau$ (oblate source). }
\begin{center}
P.E.D. / S.E.C. / P.R.P.  / $M-Q^{(1)}$ Oblate source
\begin{tabular}{ccccccccccccccc}
\hline\hline
$\tau \diagdown    q$ &$\quad$&
-0.8 &$\quad$ &
-0.5 &$\quad$ &
-0.1 &$\quad$ &
-0.05 &$\quad$ &
-0.01 &$\quad$ &
-0.005 &$\quad$ &
-0.001 
 \vspace{0.2cm} \\  \hline
2.67 &$\quad$ & V V V  &$\quad$ &   V V  V      &$\quad$ & V  V $F^*$   &$\quad$ & F V F &$\quad$ & F  V  F &$\quad$ & F V F &$\quad$ &F V F \\ \hline
2.7 &$\quad$ & V V V   &$\quad$ &  V V V &$\quad$ & V V $F^*$   &$\quad$ & F V F &$\quad$ & F V F &$\quad$ & F  F  F &$\quad$ &F  F  F \\ \hline
2.8 &$\quad$ & V V V  &$\quad$ &    V V V     &$\quad$ & V V $F^*$   &$\quad$ & F V F &$\quad$ & F  F  F &$\quad$ & F  F  F &$\quad$ &F  F  F   \\ \hline
2.9 &$\quad$ & V V V   &$\quad$ &   V V V     &$\quad$ &F V F       &$\quad$ & F V F &$\quad$ & F  F  F &$\quad$ & F  F  F &$\quad$ &F  F  F  \\ \hline
3 &$\quad$ &  V V V &$\quad$ &  V V V&$\quad$ & F V F      &$\quad$ & F  F  F &$\quad$ & F  F  F &$\quad$ & F  F  F &$\quad$ &F F  F \\ \hline
3.1 &$\quad$ & V V V &$\quad$ &  V V V&$\quad$ & F V F      &$\quad$ & F  F  F &$\quad$ & F  F  F &$\quad$ & F  F  F &$\quad$ &F  F  F  \\ \hline
3.5 &$\quad$ & V V V &$\quad$ &   V V $F^*$  &$\quad$ & F  F  F       &$\quad$ & F  F  F &$\quad$ & F  F  F &$\quad$ & F  F  F &$\quad$ &F  F  F  \\ \hline
4 &$\quad$ &  V V  $F^*$ &$\quad$ & F V $F^*$  &$\quad$ & F  F  F       &$\quad$ & F  F  F &$\quad$ & F  F  F &$\quad$ & F  F  F &$\quad$ &F  F  F   \\ \hline
4.5 &$\quad$ & F V $F^*$ &$\quad$ & F F $F^*$  &$\quad$ &  F F F   &$\quad$ & F  F  F &$\quad$ & F  F  F &$\quad$ & F  F  F &$\quad$ &F  F  F \\ \hline
5 &$\quad$ &   F F $F^*$ &$\quad$ & F F $F^*$  &$\quad$ & F  F  F       &$\quad$ & F  F  F &$\quad$ & F  F  F &$\quad$ & F  F F &$\quad$ &F  F  F \\ 
\hline\hline
\end{tabular}
\end{center}
\label{data}
\end{table}

\end{widetext}

\begin{widetext}

\begin{table}[htdp]
\caption{fulfillment (F) or violation (V) of different  criteria for good physical behaviour: Positive Energy Density (P.E.D.) ($-T_0^0>0$), Strong Energy Condition (S.E.C.) ($(-T_0^0)-T_i^i>0$) and Positive Radial Pressure (P.R.P.) ($P_{rr}=g_{11}T_1^1>0$). The symbol $^*$ over F means  that although the criterion is fulfilled, nevertheless $\partial_r P_{rr}$ changes its sign in the interval $s \in [0,1]$ within the source.
This table corresponds to the $M-Q^{(1)}$ solution for positive values of $q$ and a sequence of different values of the parameter $\tau$ (prolate source). }
\begin{center}
P.E.D. / S.E.C. / P.R.P.  / $M-Q^{(1)}$ Prolate source
\begin{tabular}{ccccccccccccccc}
\hline\hline
$\tau \diagdown    q$ &$\quad$&
0.8 &$\quad$ &
0.5 &$\quad$ &
0.1 &$\quad$ &
0.05 &$\quad$ &
0.01 &$\quad$ &
0.005 &$\quad$ &
0.001 
 \vspace{0.2cm} \\  \hline
2.67 &$\quad$ & V V $F^*$  &$\quad$ &   V V$F^*$     &$\quad$ & F V $F^*$   &$\quad$ & F V $F^*$ &$\quad$ & F V F &$\quad$ & F V F &$\quad$ &F F   F \\ \hline
2.7 &$\quad$ &V V  $F^*$  &$\quad$ &   V V  $F^*$    &$\quad$ & F V $F^*$  &$\quad$ & F V $F^*$ &$\quad$ & F F   F &$\quad$ & F F F &$\quad$ &F F F \\ \hline
2.8 &$\quad$ & V V $F^*$  &$\quad$ &   F V $F^*$    &$\quad$ & F F $F^*$   &$\quad$ & F F   $F^*$ &$\quad$ & F F F &$\quad$ & F F F &$\quad$ &F F F   \\ \hline
2.9 &$\quad$ & V V  $F^*$   &$\quad$ &  F V $F^*$      &$\quad$ &F F $F^*$      &$\quad$ & F F   $F^*$ &$\quad$ & F F F &$\quad$ & F F F &$\quad$ &F F F  \\ \hline
3 &$\quad$ &   F V $F^*$    &$\quad$& F F   $F^*$  &$\quad$ & F F $F^*$ &$\quad$ & F F $F^*$ &$\quad$ & F F F &$\quad$ & F F F &$\quad$ &F F F \\ \hline
3.1 &$\quad$ &  F V $F^*$   &$\quad$ &  F F   $F^*$      &$\quad$ & F F $F^*$ &$\quad$ & F F $F^*$ &$\quad$ & F F F &$\quad$ & F F F &$\quad$ &F F F  \\ \hline
3.5 &$\quad$ & F F $F^*$    &$\quad$ &  F F   $F^*$   &$\quad$ & F F $F^*$ &$\quad$ & F F F &$\quad$ & F F F &$\quad$ & F F F &$\quad$ &F F F  \\ \hline
4 &$\quad$ &   F F $F^*$  &$\quad$ &F F  $F^*$   &$\quad$ & F F $F^*$ &$\quad$ & F F F &$\quad$ & F F F &$\quad$ & F F F &$\quad$ &F F F   \\ \hline
4.5 &$\quad$ & F F $F^*$  &$\quad$ & F F   $F^*$   &$\quad$ & F F F       &$\quad$ & F F F &$\quad$ & F F F &$\quad$ & F F F &$\quad$ &F F F \\ \hline
5 &$\quad$ &   F F $F^*$  &$\quad$ & F F   $F^*$   &$\quad$ & F F F       &$\quad$ & F F F &$\quad$ & F F F &$\quad$ & F F F &$\quad$ &F F F \\ 
\hline\hline
\end{tabular}
\end{center}
\label{data}
\end{table}

\end{widetext}

\begin{widetext}

\begin{table}[htdp]
\caption{Fulfillment (F) or violation (V) of different  criteria for good physical behaviour: Positive Energy Density (P.E.D.) ($-T_0^0>0$), Strong Energy Condition (S.E.C.) ($(-T_0^0)-T_i^i>0$) and Positive Radial Pressure (P.R.P.) ($P_{rr}=g_{11}T_1^1>0$). The symbol $^*$ over F means  that although the criterion is fulfilled, nevertheless $\partial_r P_{rr}$ changes its sign in the interval $s \in [0,1]$ within the source.
This table corresponds to the $\gamma$ metric for negative (positive)  values of $h$, $\gamma<1$ ($\gamma>1$) and a sequence of different values of the parameter $\tau$ (prolate source). }
\begin{center}
P.E.D. / S.E.C. / P.R.P.  / $\gamma$-Metric Prolate source $\|$ $\gamma$-Metric Oblate source 
\begin{tabular}{ccccccccccccccccccccc}
\hline\hline
$\tau \diagdown    h$ &$\quad$&
-0.5 &$\quad$ &
-0.1 &$\quad$ &
-0.05 &$\quad$ &
-0.01 &$\quad$ &
-0.001 &$\|$ & 0.5 &$\quad$ &
0.1 &$\quad$ &
0.05 &$\quad$ &
0.01 &$\quad$ &
0.001 
 \vspace{0.2cm} \\  \hline
2.67 &$\quad$ & V V V  &$\quad$ &   F F $F^*$     &$\quad$ &  F F $F^*$   &$\quad$ & F F F &$\quad$ & F F F &$\|$ & V V $F^*$ &$\quad$ &V V $F^*$ &$\quad$ & F V $F^*$ &$\quad$ &F V F &$\quad$ &F V F \\ \hline
2.7 &$\quad$ & V V V  &$\quad$ &   F F $F^*$   &$\quad$ &  F F $F^*$ &$\quad$ & F F F &$\quad$ & F F F &$\|$ & V V $F^*$ &$\quad$ &V V$F^*$ &$\quad$ & F V $F^*$ &$\quad$ &F V F &$\quad$ &F F F \\ \hline
2.8 &$\quad$ & V V V  &$\quad$ &  F F $F^*$   &$\quad$ & F F $F^*$   &$\quad$ & F F F &$\quad$ & F F F &$\|$ & V V $F^*$ &$\quad$ &V V$F^*$ &$\quad$ & F V $F^*$ &$\quad$ &F F F &$\quad$ &F F F  \\ \hline
2.9 &$\quad$ & V V V   &$\quad$ &   F F $F^*$     &$\quad$ &F F $F^*$      &$\quad$ & F F F &$\quad$ & F F F &$\|$ & V V $F^*$ &$\quad$ &V V $F^*$ &$\quad$ & F V $F^*$ &$\quad$ &F F F &$\quad$ &F F F \\ \hline
3 &$\quad$ &  V V V    &$\quad$&  F F $F^*$  &$\quad$ & F F $F^*$ &$\quad$ & F F F &$\quad$ & F F F &$\|$ & V V $F^*$ &$\quad$ &V V$F^*$ &$\quad$ & F V $F^*$ &$\quad$ &F F F &$\quad$ &F F F \\ \hline
3.1 &$\quad$ &   V V V   &$\quad$ &   F F $F^*$      &$\quad$ & F F $F^*$ &$\quad$ & F F F &$\quad$ & F F F   &$\|$ & V V $F^*$ &$\quad$ &V V $F^*$ &$\quad$ & F V $F^*$ &$\quad$ &F F F &$\quad$ &F F F\\ \hline
3.5 &$\quad$ & V V V    &$\quad$ &  F F $F^*$   &$\quad$ & F F $F^*$ &$\quad$ & F F F &$\quad$ & F F F  &$\|$ & V V $F^*$ &$\quad$ &V V $F^*$ &$\quad$ & FV $F^*$ &$\quad$ &F F F &$\quad$ &F F F \\ \hline
4 &$\quad$ &   V V V   &$\quad$ & F F $F^*$  &$\quad$ & F F $F^*$       &$\quad$ & F F F &$\quad$ & F F F  &$\|$ & V V $F^*$ &$\quad$ &V V $F^*$ &$\quad$ & F F $F^*$ &$\quad$ &F F F &$\quad$ &F F F\\ \hline
4.5 &$\quad$ &  V V V   &$\quad$ &  F F $F^*$   &$\quad$ & F F $F^*$       &$\quad$ & F F F &$\quad$ & F F F  &$\|$ & V V V  &$\quad$ &V V $F^*$ &$\quad$ & F F $F^*$ &$\quad$ &F F F &$\quad$ &F F F\\ \hline
5 &$\quad$ &  V V V &$\quad$ &  F F $F^*$  &$\quad$ & F F $F^*$       &$\quad$ & F F F &$\quad$ & F F F &$\|$ & V V V &$\quad$ &V V $F^*$ &$\quad$ & F F $F^*$ &$\quad$ &F F F &$\quad$ & F F F \\ 
\hline\hline
\end{tabular}
\end{center}
\label{data}
\end{table}

\end{widetext}

\section{Acknowledgments}
This  work  was partially supported by the Spanish  Ministerio de Ciencia e
Innovaci\'on under Research Projects No.  FIS2015-65140-P (MINECO/FEDER), and the Consejer\'\i a
de
Educaci\'on of the Junta de Castilla y Le\'on under the Research Project Grupo
de Excelencia GR234.

\end{document}